\documentclass[11pt, a4paper, logo, copyright]{googledeepmind}

\pdftrailerid{redacted}

\makeatletter
\renewcommand\bibentry[1]{\nocite{#1}{\frenchspacing\@nameuse{BR@r@#1\@extra@b@citeb}}}
\makeatother

\usepackage{kantlipsum, lipsum}
\usepackage{dsfont}
\usepackage{gdm-colors}
\usepackage[utf8]{inputenc}   
\usepackage{newunicodechar}   
\usepackage{amssymb}          
\usepackage{wasysym,marvosym}

\usepackage{CJKutf8}

\usepackage[authoryear, sort&compress, round]{natbib}

\usepackage{bbding}
\usepackage[utf8]{inputenc} 
\usepackage[T1]{fontenc}    
\usepackage{hyperref}       
\usepackage{url}            
\usepackage{booktabs}       
\usepackage{nicefrac}       
\usepackage{microtype}      
\usepackage{amsmath}
\usepackage{graphicx}
\usepackage{tablefootnote}
\usepackage{multicol}
\usepackage[nameinlink]{cleveref}
\usepackage{bbm}
\usepackage{multirow}
\usepackage{soul}
\usepackage{float}
\usepackage{wrapfig}
\usepackage{blindtext}
\usepackage{tablefootnote}
\usepackage{amsfonts}
\usepackage[flushleft]{threeparttable}
\usepackage{colortbl}
\usepackage{mathtools,amssymb}
\usepackage{bm}
\usepackage{makecell}
\usepackage{caption}
\usepackage{capt-of}
\usepackage{array}
\usepackage{calc}      
\usepackage{caption}   
\usepackage{subcaption}  
\usepackage{fontawesome5}  
\usepackage{xcolor,colortbl}
\usepackage[bottom]{footmisc}
\usepackage[most]{tcolorbox}  
\definecolor{darkgray}{rgb}{0.3, 0.3, 0.3}
\definecolor{brandorange}{RGB}{234, 88, 12} 

\definecolor{tokenblue}{RGB}{37, 99, 235}   
\definecolor{demobg}{RGB}{250, 250, 250}    
\newcommand{\cmark}{\ding{51}}
\usepackage{mdframed}
\usepackage{colortbl}
\usepackage{geometry}
\usepackage{colortbl}
\usepackage{array}
\usepackage{xcolor}
\usepackage{multirow}
\usepackage{booktabs}
\usepackage{graphicx}

\newcommand{\stok}[1]{{\ttfamily\textcolor{tokenblue}{<|#1|>}}}

\newtcolorbox{demobox}[1][]{
    enhanced,
    title=#1,
    fonttitle=\bfseries\sffamily,
    coltitle=white,
    colbacktitle=black,
    colback=demobg,
    colframe=black!70,
    boxrule=0.8pt,
    arc=2pt,
    left=6pt, right=6pt, top=4pt, bottom=4pt,
    toptitle=3pt, bottomtitle=3pt,
    attach boxed title to top left={yshift=-2mm, xshift=0mm},
    boxed title style={sharp corners, boxrule=0pt}
}

\hypersetup{
    colorlinks=true,       
    linkcolor=blue,        
    urlcolor=blue,         
    citecolor=blue,        
}

\usepackage{xspace}

\newcommand{\eat}[1]{}

\crefformat{section}{\S#2#1#3}

\graphicspath{{figures/}}

\title{
    \textbf{OpenOneRec Technical Report} \\
    \vspace{0.5em}
    \scriptsize \textit{\textcolor{darkgray}{An Open Foundation Model and Benchmark to Accelerate Generative Recommendation}} 
}

\renewcommand{\today}{}

\newcommand{\themodel}{OpenOneRec\xspace}
\newcommand{\thebench}{RecIF-Bench\xspace}
\newcommand{\ourmodel}{OneRec-Foundation\xspace}

\author{
    \vspace{-1.5em}\large OneRec Team \\
    \vspace{0.5em} 
    \footnotesize 
    \begin{tabular}{r l}
        \makebox[1.8em][r]{\raisebox{-0.4em}{\includegraphics[height=1.6em]{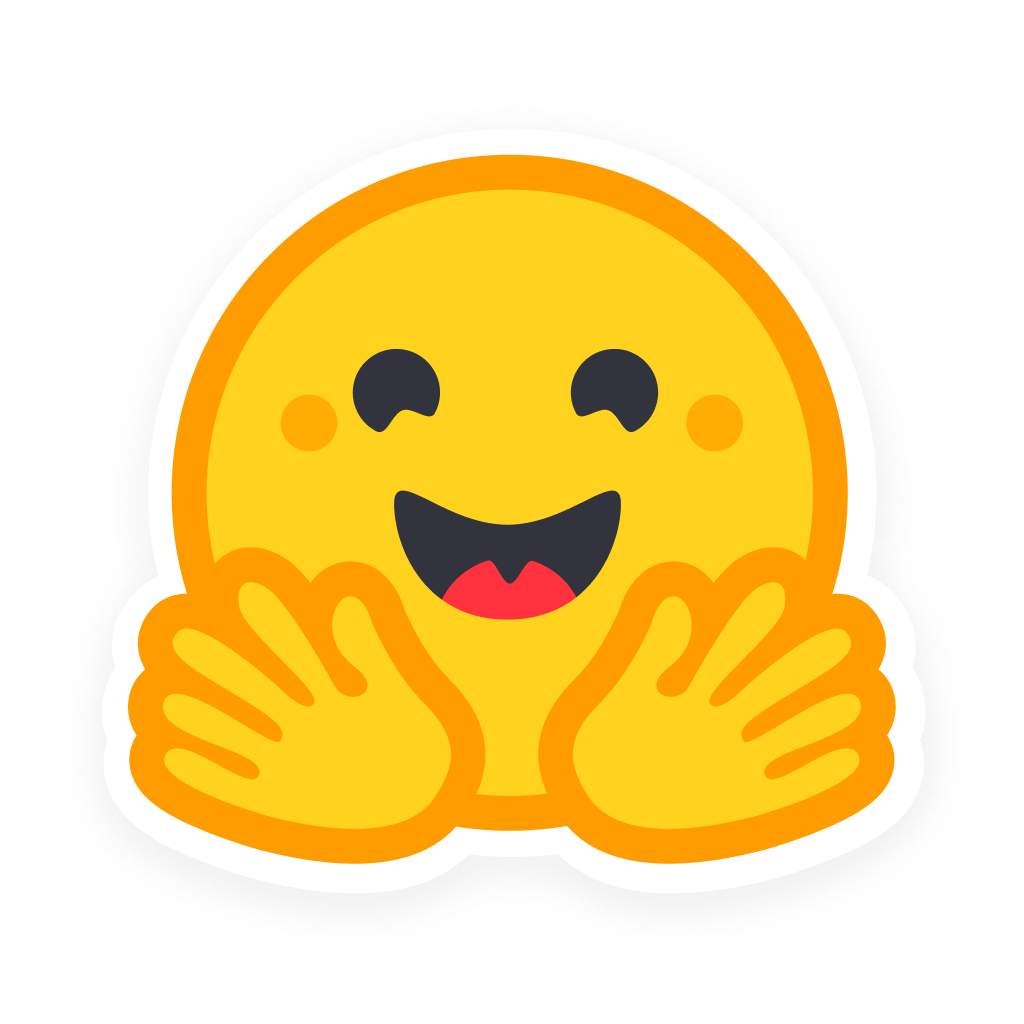}}} & \url{https://huggingface.co/OpenOneRec} \\

        \makebox[1.8em][c]{\raisebox{-0.2em}{\resizebox{!}{1.3em}{\textcolor{darkgray}{\faGithub}}}} & \url{https://github.com/Kuaishou-OneRec/OpenOneRec}
    \end{tabular}
}

\begin{document}

\begin{abstract}
While the OneRec series has successfully unified the fragmented recommendation pipeline into an end-to-end generative framework, a significant gap remains between recommendation systems and general intelligence. Constrained by isolated data, existing models struggle to leverage the massive data scaling that drives the emergent capabilities of Large Language Models (LLMs). As a result, they operate as domain specialists—proficient in pattern matching but lacking world knowledge, reasoning capabilities, and instruction following. This limitation is further compounded by the lack of a holistic benchmark to evaluate such integrated capabilities. To address this, our contributions are: 1) \thebench\ \& Open Data: We propose \thebench, a holistic benchmark covering 8 diverse tasks that thoroughly evaluate capabilities from fundamental prediction to complex reasoning. Concurrently, we release a massive training dataset comprising 96 million interactions from 160,000 users to facilitate reproducible research. 2) Framework \& Scaling: To ensure full reproducibility, we open-source our comprehensive training pipeline, encompassing data processing, co-pretraining, and post-training. Leveraging this framework, we demonstrate that recommendation capabilities can scale predictably while mitigating catastrophic forgetting of general knowledge. 3) \ourmodel: We release OneRec-Foundation (1.7B and 8B), a family of models establishing new state-of-the-art (SOTA) results across all tasks in \thebench. Furthermore, when transferred to the Amazon benchmark, our models surpass the strongest baselines with an average 26.8\% improvement in Recall@10 across 10 diverse datasets (Figure~\ref{fig:overview}).
This work marks a step towards building truly intelligent recommender systems. Nonetheless, realizing this vision presents significant technical and theoretical challenges, highlighting the need for broader research engagement in this promising direction.

\end{abstract}

\maketitle

\begin{figure}[b!]
    \centering
    \centering
        \includegraphics[width=0.95\textwidth]{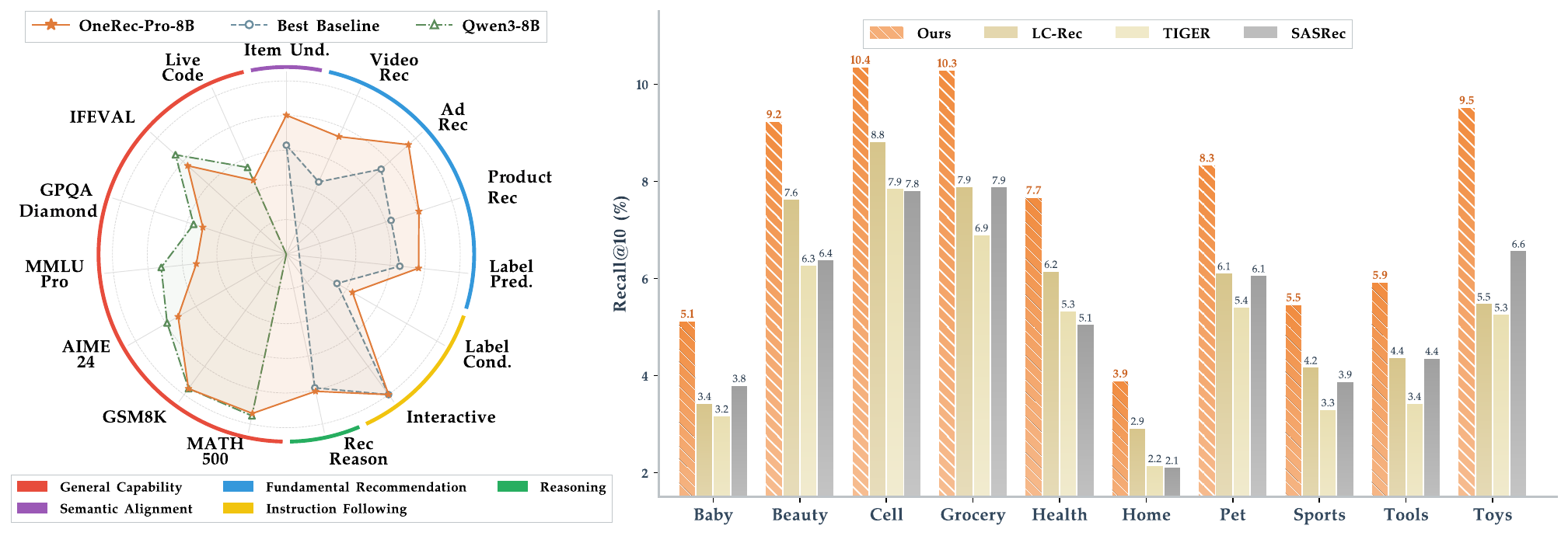}
        \caption{\textbf{Holistic Performance Overview.} \textbf{Left:} Evaluation on \thebench and general LLM benchmarks. Our model achieves SOTA performance on recommendation tasks while effectively retaining general knowledge. "Best Baseline" denotes the highest performance achieved by existing methods for each specific task. \textbf{Right:} Amazon Benchmark results. Our model demonstrates exceptional cross-domain transferability, consistently surpassing leading baselines across 10 diverse datasets.}
    \label{fig:overview}
\end{figure}

\newpage
\setcounter{tocdepth}{2} 

\tableofcontents 

\newpage
\section{Introduction}
\label{sections:1_introduction}
Recent advances in Large Language Models (LLMs) have catalyzed a paradigm shift toward Generative Recommendation \citep{geng2022recommendation, wang2023survey}. Notable efforts, such as the OneRec series \citep{deng2025onerec,zhou2025onerec,zhou2025onerecv2}, have successfully unified the multi-stage recommendation pipeline into an end-to-end generative framework, demonstrating the potential of treating user history as a context for next-item prediction. Despite these structural advancements, a significant chasm remains between current recommendation systems and general intelligence. Existing generative recommenders are often confined by isolated data silos, which sever the data scaling loops essential for the emergent capabilities seen in general LLMs \citep{kaplan2020scaling, hoffmann2022training}. Consequently, these systems function as domain-specific experts that are highly effective at collaborative pattern matching but lack the general world knowledge essential for broader intelligence.

\begin{figure}[t]
    \centering
    \includegraphics[width=1.0\linewidth]{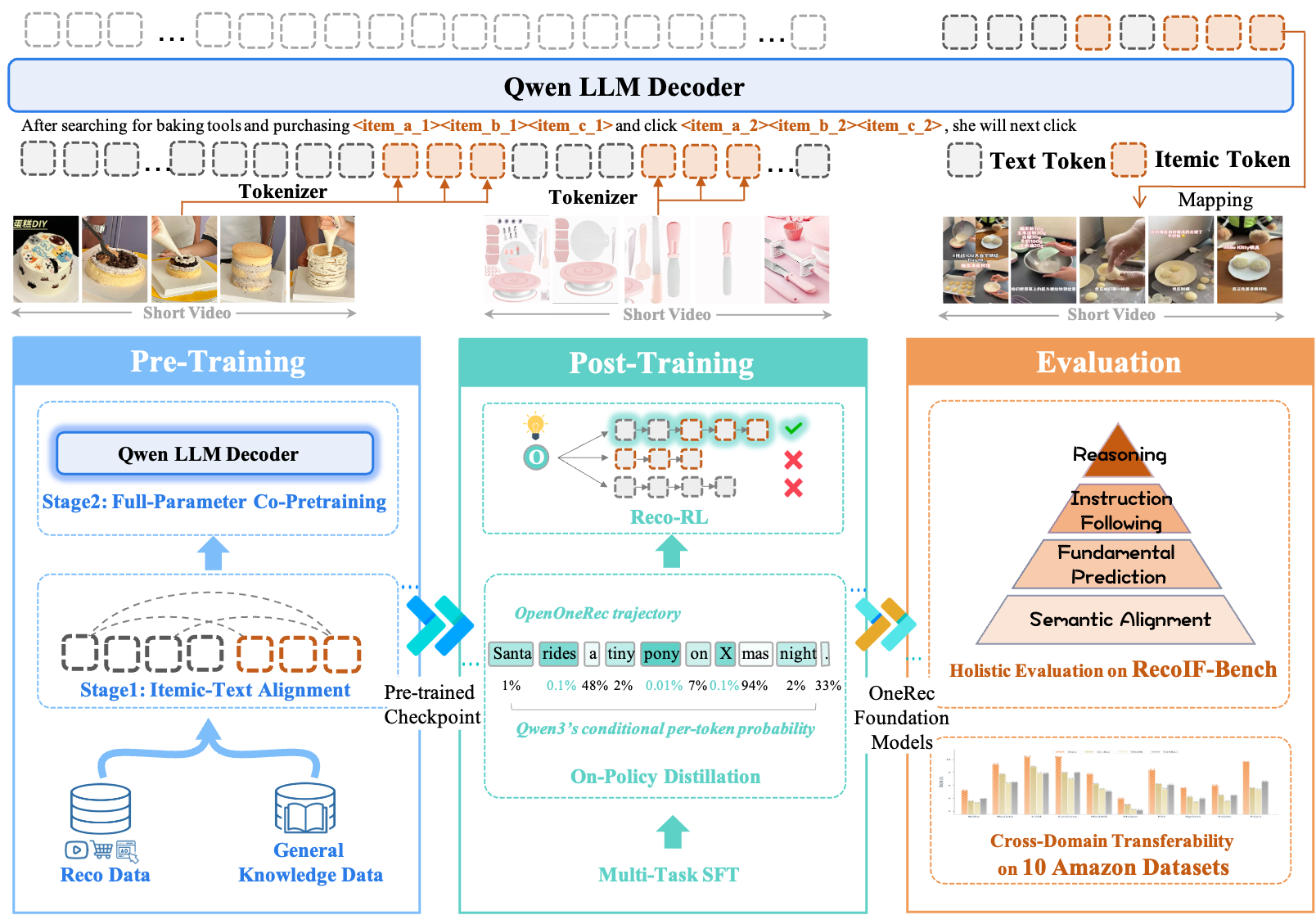}
    \caption{\textbf{The Overall Framework of \ourmodel.} (1) \textbf{Pre-Training}: Integrates collaborative signals with general semantics via Itemic-Text Alignment and mixed-domain Co-Pretraining. (2) \textbf{Post-Training}: Unlocks diverse downstream capabilities via SFT, and balances general reasoning with recommendation performance through alternating General Distillation and Rec-RL. (3) \textbf{Evaluation}: Comprehensively assesses holistic capabilities on RecIF-Bench and validates cross-domain transferability on Amazon datasets.}
    \label{framework}
    \vspace{+20pt}
\end{figure}

To bridge the semantic gap, recent studies such as LC-Rec \citep{zheng2024adapting} and OneRec-Think \citep{liu2025onerec} align discrete recommendation identifiers with the linguistic space of LLMs. However, these approaches are typically confined to a limited set of downstream tasks. Such task homogeneity often induces catastrophic forgetting, thereby compromising the backbone's inherent generalization capabilities. Furthermore, existing benchmarks remain restricted to narrow, specialized tasks, failing to evaluate the holistic capabilities essential for recommendation foundation models. In this work, we propose a unified framework (Figure \ref{framework}) that integrates scalable pre-training, hybrid post-training, and holistic evaluation. We introduce \thebench, which, to our knowledge, is the first evaluation suite designed to assess multifaceted capabilities across diverse recommendation scenarios via instruction following. To facilitate reproducibility, we open-source our complete training pipeline, including data processing, co-pretraining, and post-training protocols, along with checkpoints trained on a hundred-billion-token corpus. We summarize our main contributions as follows:
\vspace{-2pt}
\begin{itemize}[leftmargin=*, noitemsep, topsep=0pt, parsep=0pt]

\item \textbf{RecIF-Bench: A Holistic Recommendation Instruction-Following Benchmark.}
  We release \thebench, a multi-dimensional benchmark designed to rigorously assess recommender capabilities across \textbf{8 diverse tasks, ranging from fundamental recommendation to complex reasoning}. Spanning short-video, e-commerce, and online advertising, it serves as a robust testbed for quantifying the synergy between instruction following and recommendation. Furthermore, we evaluate \ourmodel on \textbf{7 widely-recognized general benchmarks} (e.g., MATH500, LiveCodeBench, GPQA-Diamond) to verify the retention of broad reasoning and coding skills. To facilitate reproducible research, we also release a comprehensive training dataset comprising \textbf{96 million interactions} from \textbf{160,000 users}.

\item \textbf{Open-Source Framework \& Validated Scaling Laws.}
To ensure full reproducibility, we open-source our complete training pipeline built upon PyTorch \citep{paszke2019pytorch} and VeRL \citep{sheng2024hybridflow}, covering data processing, co-pretraining, and post-training. We introduce a novel two-stage alignment strategy—combining \textbf{on-policy distillation and recommendation-oriented Reinforcement Learning (Rec-RL)}—to simultaneously recover general reasoning abilities and refine task-specific precision. Based on this methodology, we empirically validate scaling laws in recommendation, demonstrating predictable capability scaling while effectively mitigating the catastrophic forgetting of general knowledge.

\item \textbf{\ourmodel Model Family.}
    We release the \textbf{\ourmodel series, comprising models with 1.7B and 8B parameters}. Built on Qwen \citep{qwen3}, \ourmodel moves beyond simple fine-tuning to endow the backbone with intrinsic recommendation capabilities. The series includes \textit{Standard} versions trained on our open-source dataset and \textit{Pro} versions enhanced with a hundred-billion-token industrial corpus from Kuaishou. Empirical results show SOTA performance across all \thebench tasks. Notably, on 10 Amazon datasets, our model achieves a \textbf{substantial lead over strong baselines, securing an average 26.8\% improvement in Recall@10}, underscoring its robustness as a foundation model.

\end{itemize}

\section{Preliminary}
\label{sections:2_preliminary}
\vspace{+5pt}
In this section, we reframe the recommendation problem through the lens of generative language modeling. We introduce a modality-aligned tokenization strategy for items and formulate the recommendation task as a standard autoregressive generation problem.

\subsection{Items as Tokens: Bridging the Modality Gap}
\vspace{+5pt}
\label{sec:item_tokenization}
A fundamental challenge in applying LLMs to recommendation is the mismatch between the continuous linguistic space and the discrete item space. Representing items with detailed textual descriptions is inefficient due to excessively long contexts for user histories \citep{yang2024item} and fails to guarantee the generation of in-corpus items \citep{zheng2024adapting}. To address this, we treat items as a distinct modality, adopting \textit{Itemic Tokens} \citep{zhou2025onerec, luo2025qarm}(see Tokenizer in Figure~\ref{framework}). 

We employ RQ-Kmeans \citep{luo2025qarm} to discretize pre-trained semantic embeddings of item metadata into hierarchical discrete codes. This compresses item semantics into short, fixed-length sequences, enabling efficient long-context modeling while preserving collaborative structure. Crucially, the hierarchical nature of these tokens ensures that items with similar semantics share common prefixes, allowing the model to transfer knowledge based on token proximity—analogous to how semantic relationships are encoded in natural language tokens.

\subsection{Recommendation as Autoregressive Modeling}
\label{sec:task_formulation}
With the item modality aligned to the token space, we unify the vocabulary $\mathcal{V} = \mathcal{V}_{text} \cup \mathcal{V}_{item}$. This allows us to treat a user's interaction history not as a specialized data structure, but simply as a \textbf{long context sequence} containing item tokens, optionally interleaved with text.

We formulate tasks ranging from prediction (e.g., retrieval) to reasoning (e.g., explanation) as a unified \textbf{Next-Token Prediction} problem. Formally, given an instruction $\mathcal{I}$ and a user context $\mathcal{C}$ (comprising the interaction history and optional queries), the model maximizes the likelihood of the target response $Y$:
\begin{equation}
    \mathcal{L}(\theta) = - \sum_{t=1}^{|Y|} \log P_\theta(y_t | \mathcal{I}, \mathcal{C}, y_{<t}), 
\end{equation}
where $Y$ can be a target item's token sequence (for recommendation tasks) or a natural language sequence (e.g., item metadata, user profile, or reasoning). This formulation allows us to leverage the standard transformer architecture and training infrastructure of LLMs without any task-specific architectural modifications.

\section{\thebench: A Holistic Recommendation Instruction-Following Benchmark}
\label{sec:benchmark}

The evolution of recommender systems from specialized experts to general-purpose foundation models necessitates a paradigm shift in evaluation. Traditional benchmarks, typically confined to closed-set ranking accuracy within single domains, fall short of assessing the broader capabilities of Large Language Models (LLMs). To bridge this gap, we introduce \textbf{\thebench}, a comprehensive benchmark designed to rigorously evaluate recommendation foundation models. 

\subsection{Dataset Construction}
\label{subsec:data_construction}
\begin{figure}[t]
    \centering
    \includegraphics[width=\textwidth]{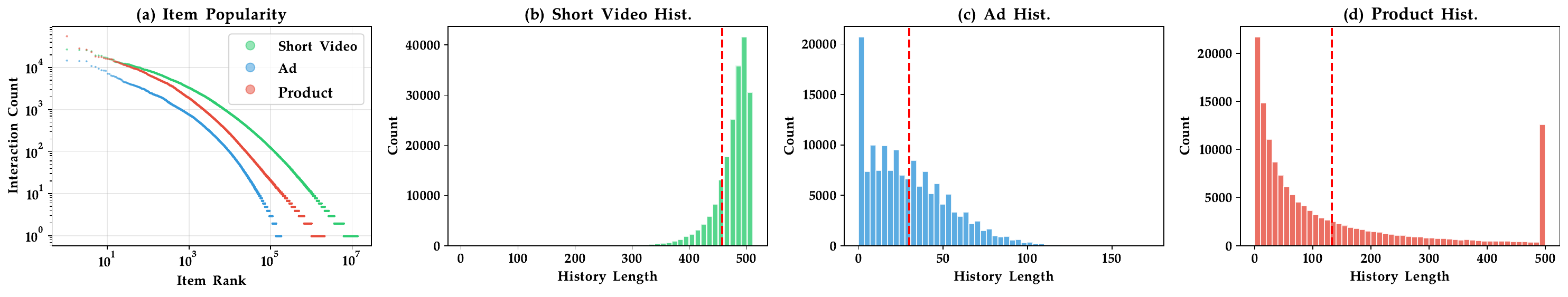}
    \caption{Data distribution analysis of \thebench. (a) Item popularity distribution (log-log scale) across domains. (b-d) Distribution of user history lengths for Short Video, Ad, and Product domains, respectively.}
    \label{fig:data_distribution}
\end{figure}

A prerequisite for a foundation recommendation model is a data substrate that encompasses diverse interaction patterns and rich semantics. \thebench aggregates approximately 120M interactions from 200K distinct users, spanning three heterogeneous industrial domains. The detailed statistics are presented in Table \ref{tab:data_stats}, and the data distributions are visualized in Figure \ref{fig:data_distribution}.

\begin{table}[h]
\centering
\caption{Statistics of \thebench. The dataset covers three distinct industrial domains with diverse interaction densities.}
\label{tab:data_stats}
  \resizebox{0.85\textwidth}{!}{
\begin{tabular}{l|r|r|r|r|r}
\toprule
\textbf{Domain} & \textbf{\# Users} & \textbf{\# Items} & \textbf{\# Interactions} & \textbf{Avg. Hist. Item} & \textbf{Avg. Tgt. Item}  \\
\midrule
\textbf{Short Video} & 195,026 & 13,107,675 & 94,443,611 & 458.1 & 8.6 \\
\textbf{Ad} & 151,259 & 177,548 & 5,341,911 & 29.9 & 5.5 \\
\textbf{Product} & 144,307 & 2,055,240 & 20,087,210 & 132.5 & 6.7 \\ \midrule
\textbf{Total} & 202,359 & 15,340,463 & 119,872,732 & 574.9 & 17.5 \\

\bottomrule
\end{tabular}
}
\end{table}

\paragraph{Multi-Domain Coverage.}
The benchmark spans three distinct domains, each capturing different user behavior patterns:
\begin{itemize}
    \item \textbf{Short Video (Content Domain)}: Short-form videos from Kuaishou, covering viewing behaviors across various APP tabs. We provide \textit{impression sequences} along with the corresponding interaction type for each impression.
    \item \textbf{Ad (Commercial Domain)}: Promotional short videos sponsored by advertisers on the Kuaishou platform, typically containing clickable redirects. We provide \textit{click sequences} recording user ad click behaviors.
    \item \textbf{Product (E-commerce Domain)}: Products listed in the Kuaishou Mall. We provide \textit{click sequences} recording user product click behaviors.
\end{itemize}

\paragraph{Rich Metadata.}
Beyond interaction logs, \thebench provides comprehensive metadata across three dimensions:
\begin{itemize}
    \item \textbf{User-side}: Each user is represented by a \textit{User Portrait}—a unified narrative that interleaves natural language descriptions with itemic tokens. This portrait weaves together demographics (gender, age), content creation history, recent searches, followed creator types, viewing preferences, comments, livestream views, purchase records, shopping cart items, local service coupons, ad exposures, and commercial intent signals.
    \item \textbf{Item-side}: Each item is associated with multimodal embeddings (4096-dim text embedding and 5-frame visual embeddings with 1152-dim per frame). Additionally, we provide dense captions for approximately 13M videos.
    \item \textbf{Interaction-side}: For each user-video pair in the exposure sequence, we record multi-label behavioral signals including like, follow, comment, effective view, and dislike.
\end{itemize}

\paragraph{Itemic Tokenization.}
To align these datasets with the generative paradigm defined in Section \ref{sec:item_tokenization}, we apply the hierarchical quantization strategy to all items in the dataset. Each item in \thebench is thus pre-tokenized into a tuple of discrete tokens $s = (c_1, c_2, ..., c_k)$, enabling direct consumption by LLM-based recommenders without further preprocessing. While we provide these pre-computed itemic tokens for convenience, their usage is \textbf{not mandatory}. Researchers can alternatively: (1) leverage the provided multimodal embeddings to train custom itemic tokenizers tailored to their specific needs, or (2) adopt traditional item IDs for conventional recommendation approaches. This flexibility ensures that \thebench accommodates diverse methodological paradigms while maintaining a consistent evaluation framework.

\paragraph{Data Splitting Strategy.}
We adopt a strict user-based splitting strategy to evaluate generalization capabilities. From the pool of 200,000 users, we randomly select 20\%  as the held-out test set. These users and their interactions are entirely excluded from the training phase, ensuring zero leakage of test user data into the training corpus. For each user, we further partition their interaction sequence temporally: interactions before a designated timestamp constitute the \textit{history} $\mathcal{H}$, while those after serve as the \textit{target} $Y$ for evaluation.

\subsection{Task Taxonomy: From Alignment to Reasoning}
\label{subsec:task_taxonomy}

\thebench organizes 8 distinct tasks into a four-layer capability hierarchy. This taxonomy allows us to pinpoint exactly where a model's capability lies on the spectrum from a traditional recommender to an intelligent agent. Table \ref{tab:task_definitions} provides a formal summary of these tasks.

\paragraph{Task Formulation.}
Following the unified formulation in Section \ref{sec:task_formulation}, we instantiate each task as a sequence-to-sequence problem $Y = \mathcal{F}(X)$. An input sample is formalized as $X = [\mathcal{I}, \mathcal{C}]$, where $\mathcal{I}$ is the task-specific instruction and $\mathcal{C}$ denotes the personalized user context. Depending on the task, $\mathcal{C}$ can take different forms: (1) the interaction history $\mathcal{H}_u = \{s_{i_1}, s_{i_2}, ..., s_{i_t}\}$ representing the user's sequential behaviors, or (2) the user portrait $\mathcal{P}_u$, a unified narrative interleaving natural language descriptions with itemic tokens. The target $Y$ can be a next item ID, a sequence of IDs, or a natural language explanation.

\begin{figure*}[t]
\centering
\includegraphics[width=0.9\textwidth]{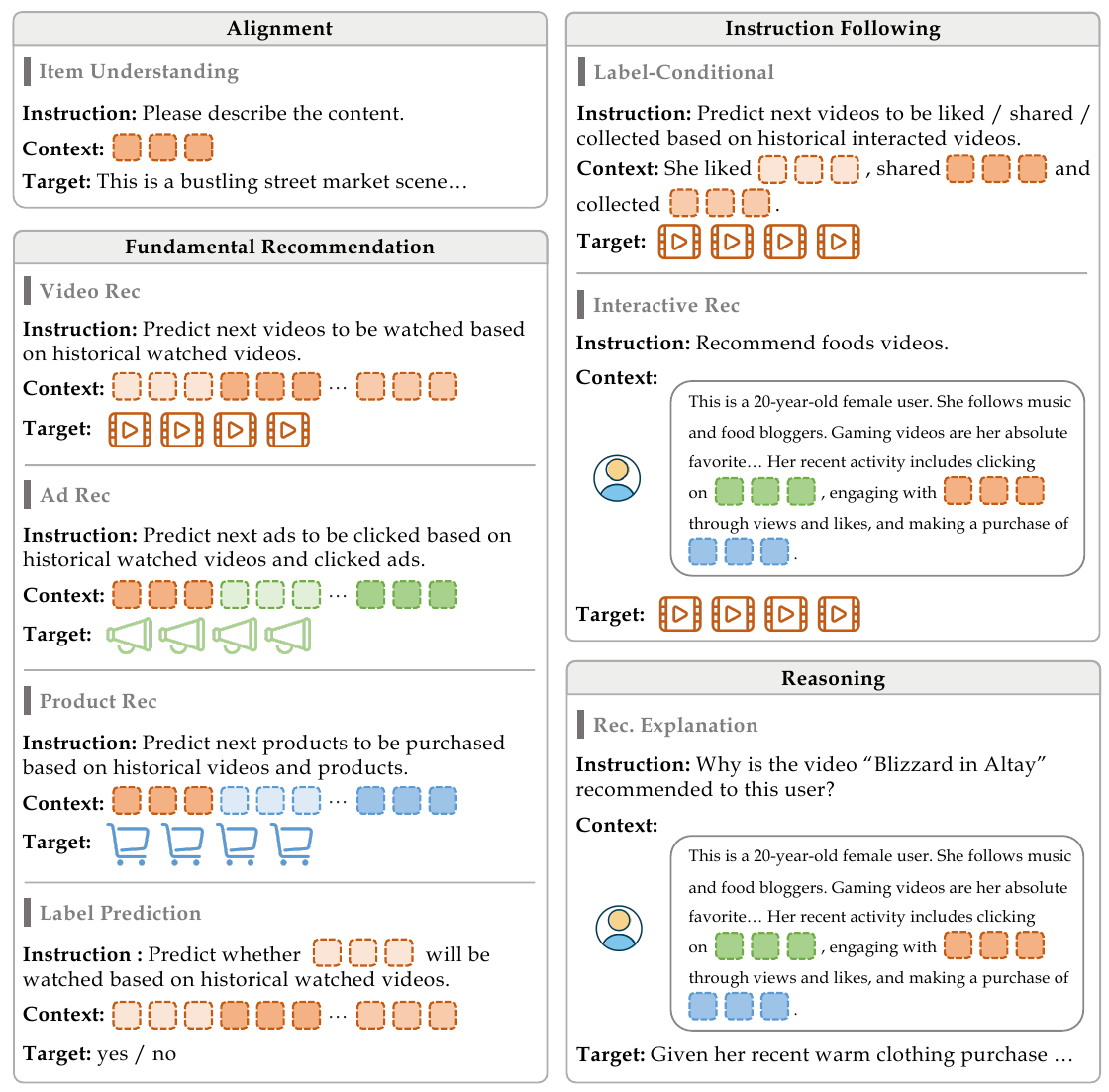}
\caption{Task Taxonomy of \thebench. We organize 8 tasks across 4 capability layers, specifying the instruction, context, and target.}
\label{fig:task_definitions}
\end{figure*}

\begin{table*}[t]

\centering

\caption{Task Taxonomy of \thebench. We formalize 8 tasks across 4 capability layers, specifying the input/output format and evaluation focus. }

\label{tab:task_definitions}

\resizebox{\textwidth}{!}{

\begin{tabular}{l|l|l|l|l}
\toprule
\textbf{Layer} & \textbf{Task} & \textbf{Input ($X$)} & \textbf{Target ($Y$)} & \textbf{Metric} \\
\midrule
\textbf{L0: Alignment}
& Item Understanding & Item $i$ & Item Description & LLM-as-Judge \\
\midrule
\multirow{4}{*}{\textbf{\makecell[l]{L1: Fundamental\\Recommendation}}}
& Short Video Rec & History $\mathcal{H}^{video}$ & Next item $i_{video}$ & Pass@1/32, Recall@32 \\
& Ad Rec & $\mathcal{H}^{video} + \mathcal{H}^{ad}$ & Next item $i_{ad}$ & Pass@1/32, Recall@32 \\
& Product Rec & $\mathcal{H}^{video} + \mathcal{H}^{product}$ & Next item $i_{product}$ & Pass@1/32, Recall@32 \\
& Label Pred. & $\mathcal{H}^{video}$ + Item $i_{video}$ & Binary (Yes/No) & AUC \\
\midrule
\multirow{2}{*}{\textbf{\makecell[l]{L2: Instruction\\Following}}}
& Interactive Rec & Portrait $\mathcal{P}$ + Query $q$ & Item $i$ user engages with & Pass@1/32, Recall@32 \\
& Label-Cond. Rec & $\mathcal{H}^{video}$ + Action $a$ & Item $i$ with action $a$ & Pass@1/32, Recall@32 \\
\midrule
\multirow{1}{*}{\textbf{L3: Reasoning}}
& Rec. Explanation & $\mathcal{P}$ + $\mathcal{H}^{video}$ + Item $i$ & Explanation & LLM-as-Judge \\
\bottomrule
\end{tabular}}

\end{table*}

\subsubsection{Layer 0: Semantic Alignment}
This foundational layer verifies whether the model has successfully bridged the modality gap between itemic tokens and natural language—a prerequisite for all downstream tasks.

\begin{itemize}
    \item \textbf{Item Understanding.}
    Given an item $i$, the model generates its textual metadata (e.g., title, caption). This inverse mapping tests whether the model aligns itemic tokens with language tokens. 
\end{itemize}

\subsubsection{Layer 1: Fundamental Recommendation}
This layer evaluates the model's core capability to capture user preferences and predict item utility across multiple domains.

\begin{itemize}
    \item \textbf{Short Video Recommendation.}
    Given the user's video viewing history $\mathcal{H}^{video}$, the model predicts the next video the user will interact with. This single-domain task serves as the canonical next-item prediction benchmark.
    \item \textbf{Ad / Product Recommendation (Cross-Domain).}
    Beyond single-domain modeling, we explore the potential of leveraging cross-domain data for more comprehensive user understanding. By integrating the user's rich content viewing history ($\mathcal{H}^{video}$) with commercial behaviors ($\mathcal{H}^{Ad}$ or $\mathcal{H}^{Product}$), the model can capture a more holistic view of user interests. Given $\mathcal{H}^{video}$ and $\mathcal{H}^{Ad}$ (or $\mathcal{H}^{Product}$), the model predicts the next ad (or product). This tests the model's capability in cross-domain interest transfer and holistic user modeling. 
    \item \textbf{Label Prediction.}
    Given the user's history $\mathcal{H}^{video}$ and a candidate item $i$, the model predicts whether the user will engage (e.g., effective view) with a binary \texttt{Yes}/\texttt{No} response. This point-wise estimation complements generative recommendation. 
\end{itemize}

\subsubsection{Layer 2: Instruction Following}
This layer assesses whether the model can adapt its predictions based on explicit natural language instructions, a key differentiator of LLM-based recommenders.

\begin{itemize}
    \item \textbf{Interactive Recommendation.}
    Given the user portrait $\mathcal{P}$ and a natural language query $q$ expressing immediate intent (e.g., "something relaxing"), the model predicts items that the user will positively engage with (e.g., click, like, or favorite) after issuing the query.
    \item \textbf{Label-Conditional Recommendation.}
    Given the user's history $\mathcal{H}^{video}$ and a target behavior label $a$ (e.g., \textit{Like}, \textit{Share}), the model predicts items the user will engage with under that specific behavior, testing fine-grained behavior modeling. 
\end{itemize}

\subsubsection{Layer 3: Reasoning}
The pinnacle of the hierarchy tests the model's ability to synthesize information and articulate its understanding in natural language.

\begin{itemize}
    \item \textbf{Recommendation Explanation.}
    Given the user portrait $\mathcal{P}$, interaction history $\mathcal{H}^{video}$, and a recommended item $s$, the model generates a natural language justification explaining why this item matches the user's profile. This tests the model's ability to reason about user-item compatibility. 
\end{itemize}

\noindent\textit{Ground Truth for L3:} Since reasoning tasks lack natural ground truth, we use Gemini-2.5-Pro with full metadata access to generate high-quality reference outputs.

\subsection{Evaluation Protocols}
\label{subsec:evaluation_metrics}
We employ a dual-metric evaluation system to cover both recommendation accuracy and generation quality.

\paragraph{Recommendation Metrics.} For recommendation tasks (Layer 1 \& 2), we use \textbf{Pass@1}, \textbf{Pass@32}, and \textbf{Recall@32}. Pass@K measures whether the ground truth item appears in the top-K generated candidates; Recall@K measures the proportion of relevant items retrieved.

\paragraph{Text Generation Metrics.} For text generation tasks (Layer 0 \& 3), we employ \textbf{LLM-as-Judge}, prompting an independent LLM to rate the generated text on dimensions such as accuracy and coherence. Details are provided in Appendix~\ref{appendix:eval_item_understanding}.

\subsection{Comparison with Existing Benchmarks}
\label{subsec:bench_comparison}

Evaluating foundation models for recommendation requires assessing capabilities beyond traditional collaborative filtering. We identify three essential dimensions:
(1) \textbf{Data Complexity}: handling multi-modal content, cross-domain transfer, long interaction sequences, and diverse user behaviors;
(2) \textbf{Task Generalization}: executing multiple recommendation tasks within a unified framework;
(3) \textbf{Instruction Following \& Reasoning}: processing interleaved inputs (text and itemic tokens) and generating explainable recommendations.

Existing benchmarks address these dimensions only partially. As shown in Table~\ref{tab:bench_comparison}, PixelRec and NineRec focus on multi-modal features but lack multi-task and reasoning capabilities. KuaiSAR supports multi-task and multi-behavior settings but omits multi-modal and cross-domain scenarios. Critically, no existing benchmark evaluates \textit{Interleaved Data} processing or \textit{Recommendation Explanation}---two capabilities essential for instruction-tuned foundation models. \thebench is the first benchmark to cover all seven dimensions, providing a comprehensive testbed for next-generation recommendation foundation models.

\begin{table}[ht]
\centering
\caption{Feature Comparison of Recommendation Benchmarks. \thebench is the only benchmark that simultaneously supports all seven dimensions. Specifically, it uniquely integrates Multi-Behavioral interactions, Interleaved Data(combining text and itemic tokens), and Recommendation Explanation to evaluate foundation models' instruction-following ability.}
\label{tab:bench_comparison}
\small
\setlength{\tabcolsep}{2pt}
\begin{tabular}{l|ccccccc}
\toprule
\textbf{Benchmark} & \makecell{Long-Seq.} & \makecell{Multi-Task} & \makecell{Multi-Modal} & \makecell{Cross-Domain} & \makecell{Multi-Behav.} & \makecell{Interleaved Data} & \makecell{Reco. Exp.} \\
\midrule
PixelRec          & \cmark &        & \cmark &        &        &        &        \\
KuaiSAR           & \cmark & \cmark &        &        & \cmark &        &        \\
NineRec           &        &        & \cmark & \cmark &        &        &        \\
Yelp              &        &        & \cmark &        &        &        &        \\
Amazon            &        &        & \cmark & \cmark &        &        &        \\
\textbf{\thebench} & \cmark & \cmark & \cmark & \cmark & \cmark & \cmark & \cmark \\
\bottomrule
\end{tabular}
\end{table}

\subsection{Sanity Check: General Intelligence}
\label{subsec:general_eval}
To assess whether recommendation-focused models retain general intelligence capabilities, we include a sanity check suite covering four categories:
(1) \textbf{Math \& Text Reasoning}: MATH-500, GSM8K, and AIME'24;
(2) \textbf{General Tasks}: MMLU-Pro and GPQA-Diamond;
(3) \textbf{Alignment}: IFEVAL (strict prompt);
(4) \textbf{Coding}: LiveCodeBench v5.
A successful foundation model should maintain competitive performance on these benchmarks while mastering recommendation tasks.


\section{Pre-Training}
\label{sections:4_pre_training}


In this section, we detail the construction of our pre-training data, followed by the specific training strategies and an empirical analysis of scaling laws in recommendation.

\subsection{Pre-training Data}


\subsubsection{Recommendation Corpora}
\label{subsec:recommendation_data}

We derive recommendation data from anonymized user logs within Kuaishou, including user-side, item-side and interaction-side raw metadata as detailed in \ref{subsec:data_construction}. To bridge the modality gap between discrete IDs and natural language, we first extract semantic embeddings from video text descriptions utilizing the Qwen3-8B-Embedding \citep{qwen3embedding} (we also provide multimodal embeddings to facilitate future research). These embeddings are subsequently quantized into hierarchical item tokens via RQ-Kmeans \citep{luo2025qarm}. 
Specifically, we employ a three-layer quantization scheme with a codebook size of 8192 per layer. Each item $i$ is thus mapped to a tuple of hierarchical codes $S_i = (c_1, c_2, c_3)$, which is then flattened into a token sequence wrapped by special tokens: 
\begin{center}
\textbf{\texttt{<|item\_begin|><item\_a\_5028><item\_b\_6733><item\_c\_2559><|item\_end|>}}
\end{center}

To enhance the model's recommendation-domain capability mentioned in \ref{subsec:task_taxonomy}, we organize these raw metadata into three types of data: 
\begin{itemize}
    \item \textbf{Itemic Dense Caption Data.} To build an initial perception of itemic tokens, we introduce the \textit{Itemic Dense Caption} data. Given an item represented as itemic tokens, the model is trained to generate a corresponding natural-language caption. This task facilitates a semantic bridge between the abstract, discrete item representations and their rich textual descriptions. Each training sample is formatted as a structured sequence, pairing the item tokens with a target caption.

    \item \textbf{Sequential User Behavior Data.} To enhance the model's Fundamental Prediction capability, we utilize \textit{Sequential User Behavior} data as the core training corpus. This data captures the chronological dynamics of user-item interactions, including views, likes, and shares. By training the model to perform next-item prediction within these long-term sequences, we enable it to internalize fundamental collaborative filtering signals and temporal patterns. This ensures the model excels at the primary recommendation task: accurately predicting a user’s future interest based on their historical behavioral trajectory.
    
    \item \textbf{Interleaved User Persona Grounding Data.} To facilitate deep semantic grounding of the quantized space, we construct the \textit{Interleaved User Persona Grounding} data. We construct narrative-style user portraits $\mathcal{P}_u$ by interleaving discrete item representations with heterogeneous user metadata, such as static attributes (e.g., age, gender), active search behaviors (e.g., recent search queries), interactive sequences (represented as itemic token sequences from user interaction history), and summarized user interests (e.g., content creation history, followed creator types, consumption preferences). This interleaved format enables the model to learn rich semantic associations between user characteristics and behavioral patterns, facilitating a deeper understanding of user preferences and item relevance beyond surface-level sequential patterns.
\end{itemize}

The recommendation-domain datasets are derived from the raw metadata introduced in \ref{subsec:data_construction}, which has been processed by a strict user-based split to avoid data leakage. The primary training corpus encompasses about 160k users, 13 million item captions, and their corresponding interactions. For the OneRec-Pro variant, we scale this to $\sim$20 million users and 98 million item captions. To provide a clearer intuition of our pre-training corpus, we provide representative data samples in Appendix~\ref{appendix:samples_rec_domain_pre_training}. These demos show the structured formats of our training instances and illustrate the seamless integration of hierarchical itemic tokens with textual metadata. All these data and processing scripts will be released in our GitHub repository~\footnote{https://github.com/Kuaishou-OneRec/OpenOneRec} to facilitate reproducibility.


\subsubsection{General-Domain Corpora}

While injecting specialized recommendation knowledge is crucial for domain specialization, the substantial distributional shift between recommendation-domain data and the base pretrained model's original corpus often leads to catastrophic forgetting. To mitigate this degradation and ensure a stable transition of model parameters, we adopt a data-mixing strategy that co-trains the model on recommendation-domain samples alongside high-quality general-domain text corpus~\citep{karimi2025nemotroncc, basant2025nvidia, moshkov2025aimo2, ahmad2025opencodereasoning, xu2025kodcode, chen2024huatuogpto1medicalcomplexreasoning, bespoke_stratos,glaive_ai_reasoning_v1_2025,mxode_chinese_reasoning_2024}.

This general-domain text corpus contains multiple languages (including Chinese, English, and others) and various domains, mainly focusing on Coding, STEM(Science, Technology, Engineering, and Mathematics) and Medical. Crucially, to keep and further enhance the model's reasoning capability, we prioritize reasoning-intensive data, including mathematical derivations, logical puzzles, and code-centric corpora. All these data can be downloaded from the Hugging Face dataset repository~\footnote{https://huggingface.co/datasets}. For convenience, we also provide detailed data composition and download links in Appendix~\ref{appendix:data_composition_and_token_budgets}.

Specifically, we employ the MinHash algorithm \citep{broder1997resemblance} to conduct efficient fuzzy deduplication, filtering out any general-domain samples that exhibit high similarity to the evaluation benchmarks. This process ensures that the model's performance reflects genuine generalization, thereby maintaining the reliability of our experimental results.
    
\subsection{Training Recipe}

In terms of data composition, we blend recommendation-domain metadata with general-domain text at predefined ratios. Specifically, we develop two model variants based on the scale of the training corpus: \textbf{OneRec} and \textbf{OneRec-Pro}. The standard variant is trained exclusively on our publicly released dataset, encompassing 32B tokens across 41.3 million samples, thereby establishing a reproducible baseline for the community. In contrast, the Pro variant leverages an extensive in-house corpus with broader user coverage, totaling 130B tokens and 179.1 million samples to achieve enhanced robustness. A comprehensive breakdown of the token composition and mixing strategies is provided in Appendix~\ref{appendix:data_composition_and_token_budgets}.

For both variants, we initialize our model using the Qwen3 backbone \citep{qwen3}, inheriting its full parameters after post-training for instruction following and reinforcement learning. The architecture remains strictly consistent with the original Qwen3 to preserve its foundational linguistic and reasoning capabilities. Building on this architecture, our pre-training methodology is structured into two distinct stages, as illustrated in the pre-training phase of Figure~\ref{framework}. 

\paragraph{Stage 1: Itemic-Text Alignment.} The first stage focuses on establishing a preliminary alignment between the itemic tokens and text tokens space. We first expand the vocabulary by appending these itemic special tokens to the original Qwen3 tokenizer. The embedding parameters for these itemic tokens are initialized from a multivariate normal distribution parameterized by the mean and covariance of the existing embeddings. During this stage, only the embedding parameters corresponding to itemic tokens are trainable, while all other model's parameters are frozen. Note that in Qwen3, smaller models (e.g., 0.6B, 1.7B, 4B) employ tied embeddings where the embedding and output projection layers share parameters, while larger models (e.g., 8B and above) have independent output projection parameters. For larger models, the output projection parameters corresponding to itemic tokens are also trainable, ensuring proper alignment in the output space. 



\paragraph{Stage 2: Full-Parameter Co-Pretraining.} In the second stage, we unfreeze all model parameters and conduct full-parameter pre-training to inject recommendation knowledge into the model. This stage aims to enable the model to capture complex patterns in user behavior, item semantics, and user-item interactions while preserving the general world knowledge inherited from the original Qwen3 model. To prevent catastrophic forgetting, we maintain a considerable proportion of general-domain knowledge data throughout this stage. 


\paragraph{Training Recipe.} We use the AdamW optimizer with $\beta_1 = 0.9$, $\beta_2 = 0.95$, and weight decay of $0.1$. The learning rate follows a cosine decay schedule with a linear warmup phase, where the peak learning rate is set to $1 \times 10^{-3}$ for Stage 1 and $1 \times 10^{-4}$ for Stage 2, and the minimum learning rate is set to $1 \times 10^{-4}$ and $2 \times 10^{-5}$. The warmup duration spans the first $10\%$ of training steps. To accommodate the long sequential nature of user behavior data, we set the maximum context length to $32$K tokens, enabling the model to process extended user interaction histories and complex recommendation scenarios. This extended context window is crucial for capturing long-term user preferences and understanding intricate patterns in sequential recommendation tasks.

\begin{figure}[htbp]
    \centering
    \begin{subfigure}[b]{0.49\textwidth}
        \centering
        \includegraphics[width=\textwidth]{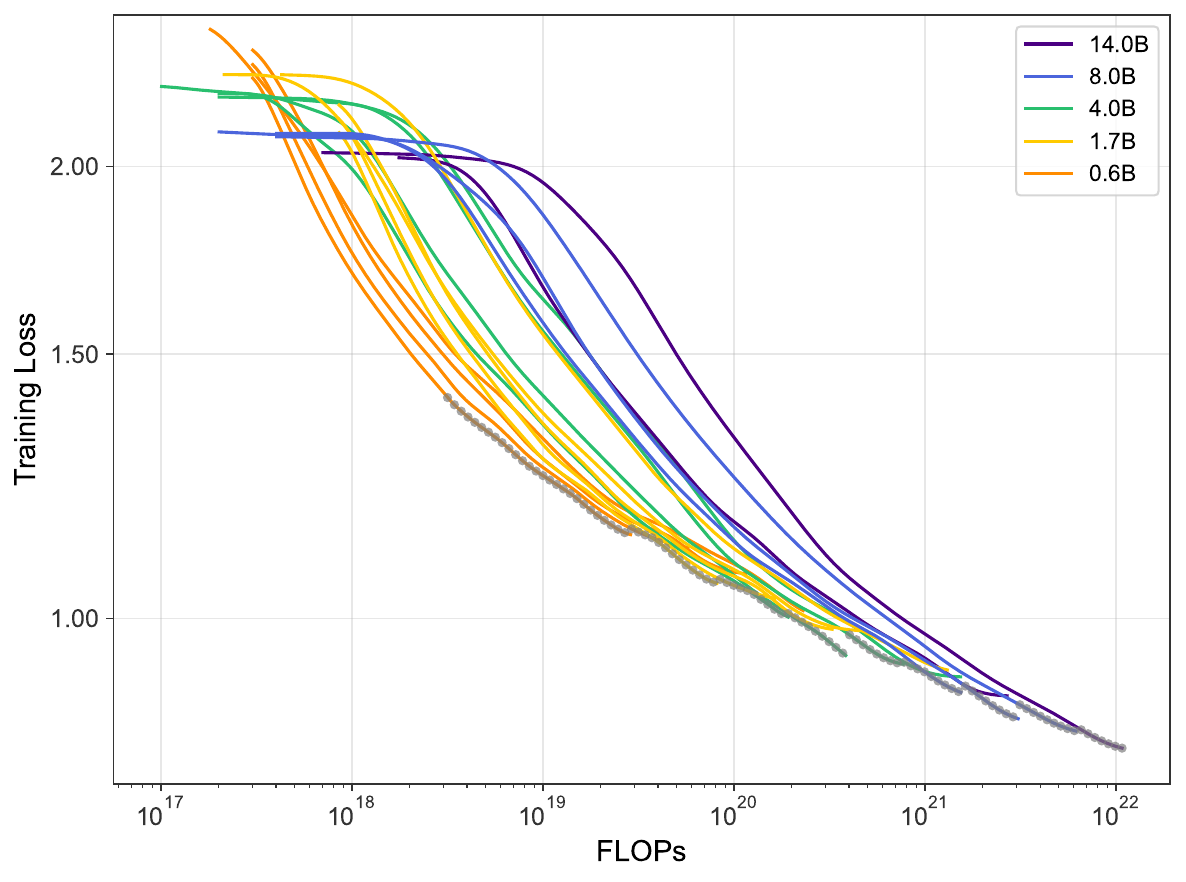}
        \label{fig:rec_loss}
    \end{subfigure}
    \hfill
    \begin{subfigure}[b]{0.49\textwidth}
        \centering
        \includegraphics[width=\textwidth]{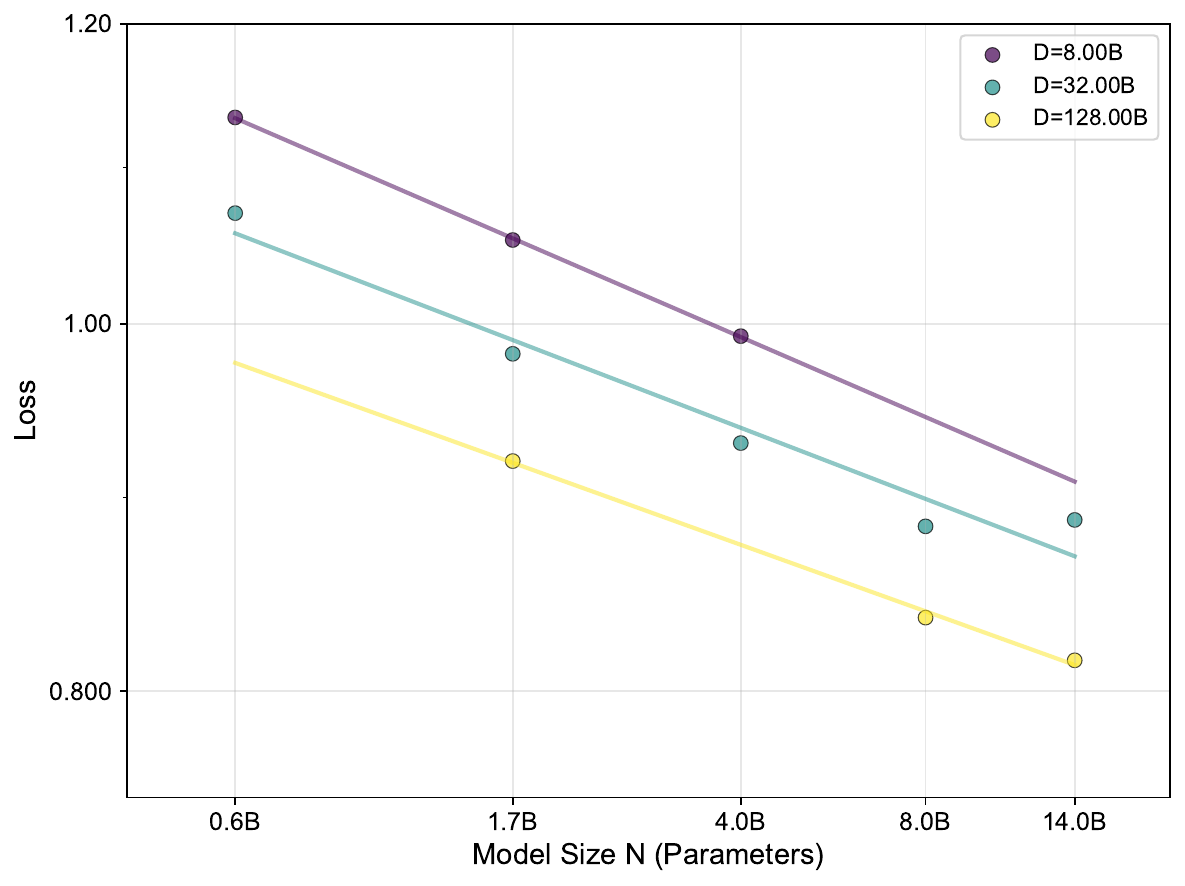}
        \label{fig:rec_loss_trends}
    \end{subfigure}
    \\[0.5em]
    \begin{subfigure}[b]{0.49\textwidth}
        \centering
        \includegraphics[width=\textwidth]{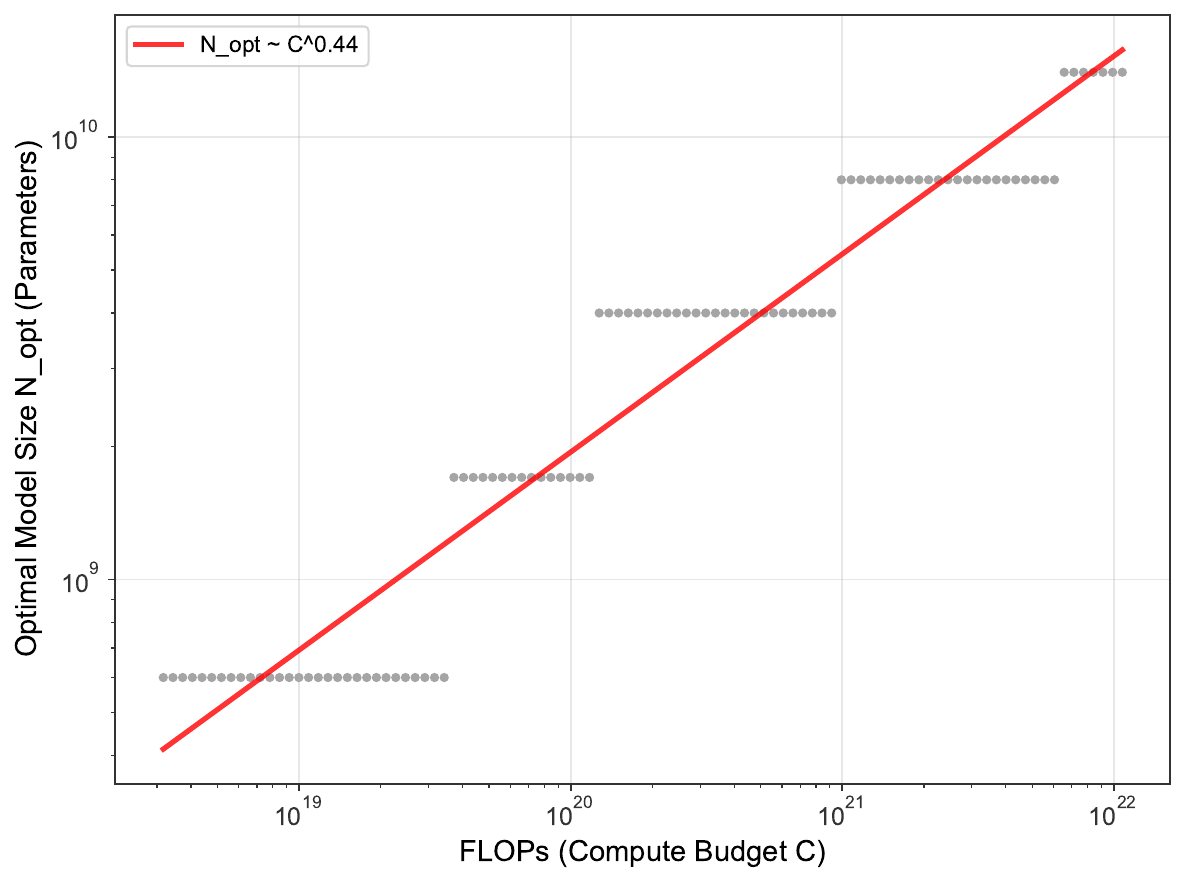}
        \label{fig:rec_params}
    \end{subfigure}
    \hfill
    \begin{subfigure}[b]{0.49\textwidth}
        \centering
        \includegraphics[width=\textwidth]{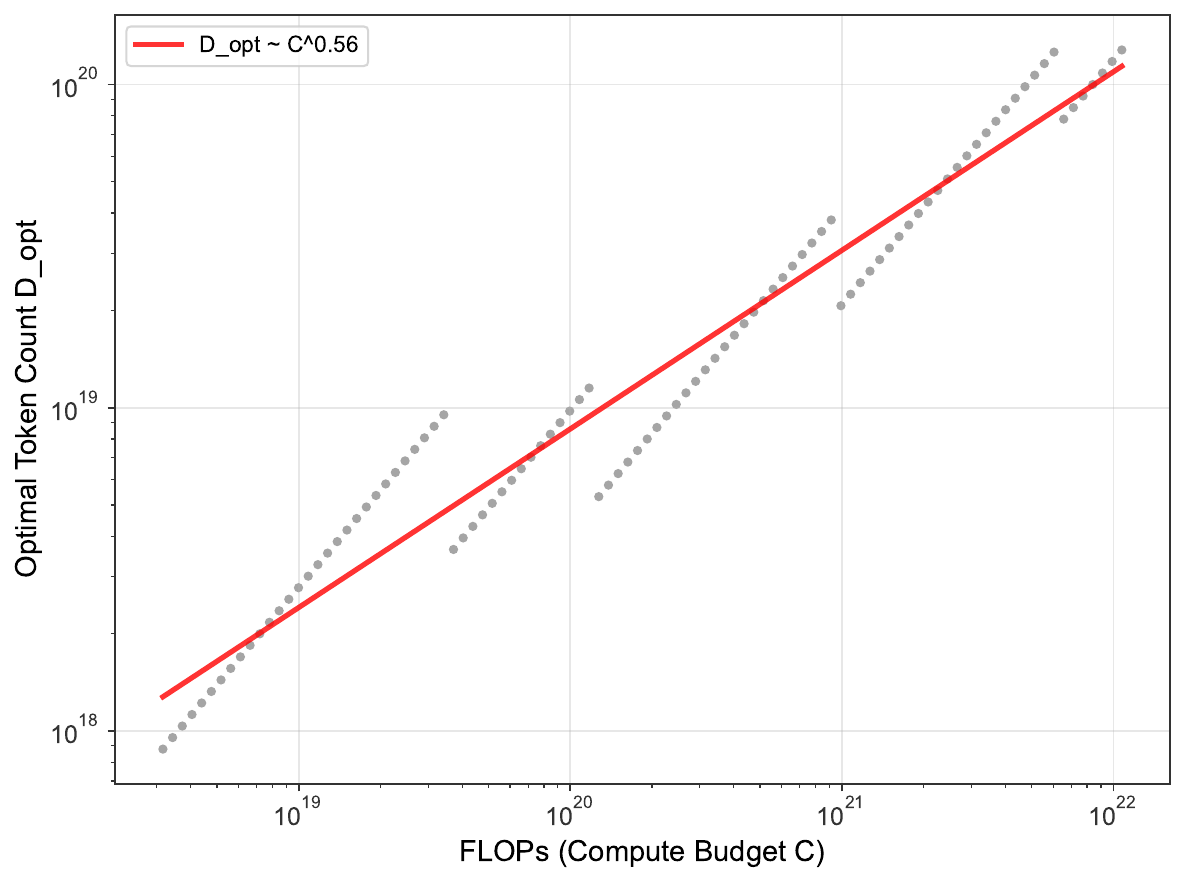}
        \label{fig:rec_tokens}
    \end{subfigure}

    \caption{\textbf{Scaling laws on recommendation-domain data.} \textbf{Top-left:} training loss vs. FLOPs for different model sizes, and the smooth convex envelope shown in grey. \textbf{Top-right:} log-log plot of training loss vs. model size under fixed token budgets. \textbf{Bottom-left:} compute-optimal model size $N_{\text{opt}} \propto C^{0.44}$ as a function of compute budget $C$. \textbf{Bottom-right:} compute-optimal token budget $D_{\text{opt}} \propto C^{0.56}$ as a function of $C$.}
    \label{fig:scaling_laws_rec}
\end{figure}

\subsection{Scaling Laws in Recommendation}

To determine the optimal allocation of compute budget $C$ between model parameters $N$ and training tokens $D$, we follow the rigorous methodology established by \citet{hoffmann2022training}. Utilizing the Qwen3 architecture as our backbone, we evaluate scaling behaviors across a parameter spectrum of $N \in \{0.6, 1.7, 4, 8, 14\} \times 10^9$. For the scaling analysis, we vary the token budget $D$ in Stage 2 of pretraining, modulating the cosine learning rate schedule to match the training horizon for each run, as recommended to avoid sub-optimal convergence. The training compute budget is estimated using the standard approximation $C \approx 6ND$.

We define the \textit{compute-optimal frontier} by constructing the convex hull of the final training loss across all model sizes. By interpolating the minimum loss $L_{\text{min}}(C)$ on a continuous FLOPs grid, we extract the optimal model size $N_{\text{opt}}$ and token count $D_{\text{opt}}$ for a given budget. These optimal trajectories are fitted to power-law scaling relations:
\begin{equation}
    N_{\text{opt}} \propto C^{a}, \quad D_{\text{opt}} \propto C^{b}
\end{equation}


\paragraph{Scaling Laws on Recommendation Data.} 
Figure~\ref{fig:scaling_laws_rec} illustrates the recommendation-domain loss envelope. We observe a smooth, convex frontier, confirming that scaling behavior in recommendation follows predictable power laws analogous to natural language. However, our empirical fit yields the scaling exponents:
\begin{equation}
    N_{\text{opt}} \propto C^{0.44}, \quad D_{\text{opt}} \propto C^{0.56}
\end{equation}
This allocation deviates significantly from the Chinchilla scaling laws for general text, which imply an equiproportional split ($a \approx 0.5, b \approx 0.5$). In contrast, our results indicate a \textbf{data-intensive scaling regime} ($b > a$), suggesting that in the recommendation domain, optimal compute allocation requires scaling training data volume more aggressively than model parameters as the budget increases.

\paragraph{Parametric Fit and Interpretation.} 
To elucidate the drivers of this behavior, we model the final loss $L(N, D)$ using the parametric function proposed by \citet{hoffmann2022training}:
\begin{equation}
    L(N, D) = E + \frac{A}{N^{\alpha}} + \frac{B}{D^{\beta}}
    \label{eq:loss_formula}
\end{equation}
where $E$ represents the irreducible entropy of the data distribution, and the terms $\frac{A}{N^{\alpha}}$ and $\frac{B}{D^{\beta}}$ capture the deviations due to finite model capacity and finite data size, respectively. Fitting this formulation to our experimental data yields:
\begin{equation}
    L(N, D) = 0.4232 + \frac{502.32}{N^{0.3325}} + \frac{7.02}{D^{0.1865}}
    \label{eq:fit_loss_formula}
\end{equation}
Based on the theoretical relationship derived in \citet{hoffmann2022training}, the optimal scaling exponents are governed by the ratio of these decay rates: $a = \frac{\beta}{\alpha+\beta}$ and $b = \frac{\alpha}{\alpha+\beta}$. We derive three critical insights from these coefficients:

\begin{itemize}
    \item \textbf{Data-Hungry Scaling ($\alpha > \beta$):} Our derived model capacity exponent ($\alpha \approx 0.33$) is consistent with general LLM literature, but the data exponent ($\beta \approx 0.19$) is notably lower than typical text-domain values ($\beta_{\text{text}} \approx 0.28$). Since $\alpha > \beta$, it mathematically necessitates that $b > 0.5$. This confirms that because the returns on data quantity diminish more rapidly (lower $\beta$), maintaining optimality requires scaling data volume more aggressively than model size.
    
    \item \textbf{Impact of Warm-Starting (High $A$, Low $B$):} 
    We observe a striking imbalance between the scaling coefficients $A$ ($502.32$) and $B$ ($7.02$). In typical training-from-scratch scenarios, $A$ and $B$ are of comparable magnitude. We attribute the extremely low $B$ value to the efficacy of transfer learning from the Qwen3 backbone, where robust pre-trained linguistic and reasoning capabilities lower the entropy of the initial data distribution.  Conversely, the inflated coefficient $A$ reflects a conflation of model capacity and pre-training quality: as larger variants are typically trained with more data, their downstream performance gains are statistically captured into $A$ during the fitting process. Notwithstanding this imbalance, our formula aligns well with our empirical observations, reflecting the scaling trajectory between $L$ and the variables $(N, D)$ across all evaluated configurations.


    \item \textbf{Low Entropy of Recommendation Tasks (Low $E$):} The estimated irreducible loss floor $E=0.42$ is substantially lower than that of natural text ($E \approx 1.69$). This suggests that our recommendation tasks—enriched with structured features such as Itemic Dense Captions—possess lower inherent entropy than open-ended text generation, allowing the model to approach a more deterministic state. Consequently, this underscores the critical need for curating recommendation corpora with greater diversity and higher quality, thereby expanding the information manifold to prevent trivial saturation and foster robust generalization.
\end{itemize}

Formula~\ref{eq:fit_loss_formula} provides an initial exploration of the scaling law in recommendation. We acknowledge that the limited experimental scale may introduce some fitting noise. Furthermore, since our models are not trained from scratch, the formula does not fully separate the scaling contribution of the pre-trained backbone. Future work will focus on expanding the experimental scale and design refined formula that better integrate the influence of pre-trained knowledge.

\section{Post-Training}
\label{sections:5_post_training}
\begin{figure}[h]
    \centering
    \includegraphics[width=\textwidth]{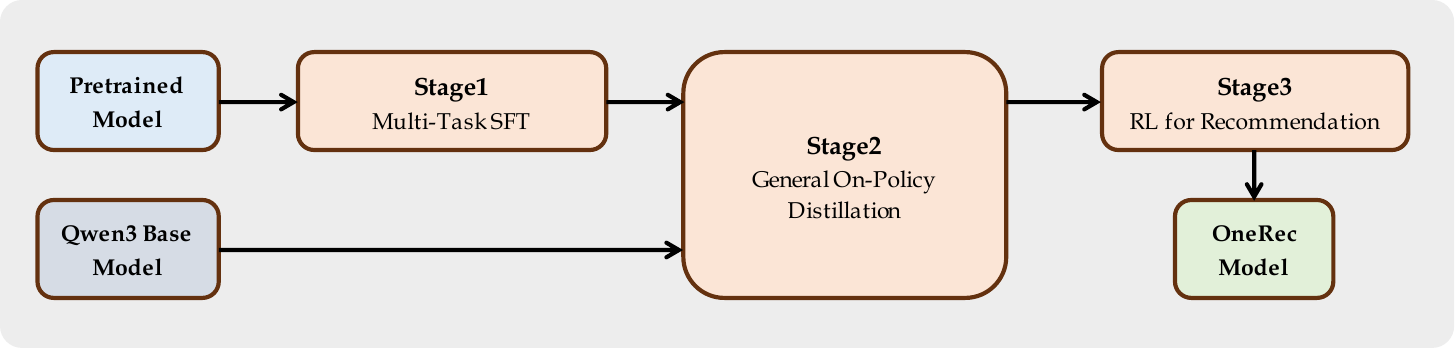}
    \caption{Post-training pipeline of the OneRec series models.}
    \label{fig:onerec-post-train-pipeline}
\end{figure}

After the pre-training process, the model has already learned to align the itemic tokens with the text tokens space, and encoded the collaborative filtering signals into the pretrained models. However, we observed that the pre-trained model exhibits a certain degree of degradation in instruction-following and reasoning capabilities, which is still not capable of complex recommendation tasks. 

The post-training is designed to enhance the recommendation capabilities and restore the general task capabilities of the pre-trained model. We adopt three post-training stages: \textbf{Multi-task Supervised Fine-tuning}, \textbf{On-policy Distillation for General Capability}, and \textbf{Reinforcement Learning for Recommendation}. Distinct from the pre-training phase, for this stage, we employ identical training data and equivalent post-training strategies for both the \textbf{OneRec} and \textbf{OneRec-Pro} variants. An overview of these stages is presented in Figure \ref{fig:onerec-post-train-pipeline}.

\subsection{Multi-task Supervised Fine-tuning}
\label{subsec:multi-task-sft}


This stage is designed to restore and enhance the model's foundational instruction-following and reasoning capabilities across both general and recommendation domains, providing a robust base for subsequent on-policy distillation and reinforcement learning.

We begin by designing a suite of complex instruction-response pairs that simulate real-world recommendation trajectories for post-training.  Consistent with the pre-training stage, these conversational datasets are synthesized based on the cleaned metadata from 160K users introduced in Section \ref{subsec:data_construction}. This ensures that the instruction-tuning corpus remains strictly partitioned from the evaluation benchmarks. Besides, to prevent the degradation of general intelligence, we also incorporate high-quality open-source general-domain datasets focusing on instruction-following and complex reasoning\citep{moshkov2025aimo2, slam-distillation-from-r1, li2025infinityinstructscalinginstruction, ahmad2025opencodereasoning, xu2025kodcode, su2025expanding, deepmath}. 


We curate a specialized SFT corpus by blending these general-domain reasoning samples with the aforementioned recommendation-specific tasks, and the detail is provided in Appendix~\ref{appendix:data_composition_for_sft}. All instances are organized into a conversational format and serialized using the Qwen3 chat template \citep{qwen3}. We fine-tune the pre-trained model on this unified dataset using a training recipe consistent with the pre-training phase, but with a reduced learning rate (from $2 \times 10^{-5}$ to $5 \times 10^{-6}$). Empirical observations suggest that this stage successfully resuscitates the model's instruction-following capabilities. Notably, we find that the reasoning ability acquired from general-domain data cross-fertilizes with recommendation tasks: the model frequently generates coherent reasoning trajectories for complex recommendation queries, even though such behaviors were not explicitly supervised in the recommendation samples.

\subsection{On-policy Distillation for General Capability}

While previous stages have successfully restored the basic capabilities of instruction-following and thinking, a persistent capability gap in general-domain reasoning is observed, likely due to the distributional shift and the inherent sensitivity of RL-initialized backbones. To address this, we design an \textbf{on-policy distillation} strategy on general tasks.

\paragraph{On-Policy Distillation via Policy Gradient.} 
Unlike traditional off-policy distillation, where the student model learns a teacher's distribution on a static, pre-generated dataset, on-policy distillation~\citep{agarwal2024policy} involves the student model generating its own trajectories, which are subsequently evaluated and supervised by the teacher.


\begin{figure}[t]
    \centering
    \includegraphics[width=0.8\linewidth]{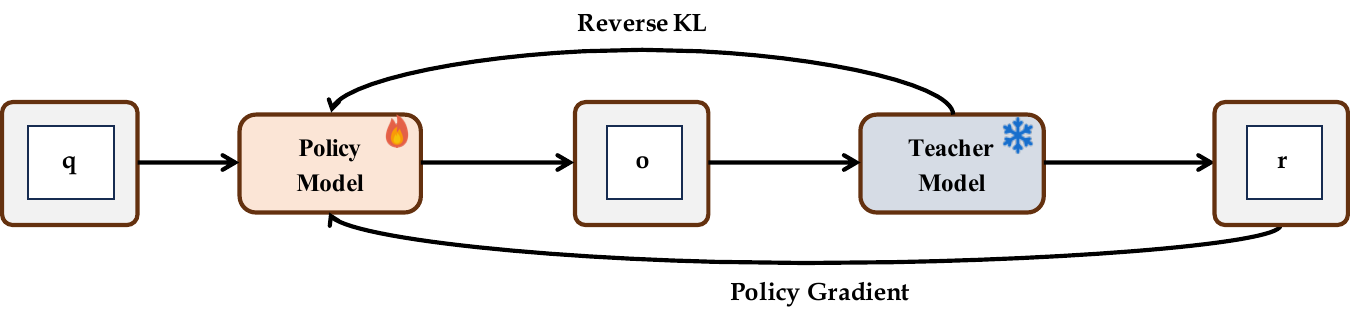}
    \caption{\textbf{The pipeline of On-policy Distillation via Policy Gradient.} The student model (Policy Model) samples a trajectory $o$ for the given prompt $q$, while the Teacher Model provides feedback through Reverse KL divergence as reward $r$. The Policy Model is iteratively optimized using policy gradient methods based on reward.}
    \label{fig:onpolicy-distillation-pipeline-single}
\end{figure}

For simplicity and effectiveness, we employ per-token reverse KL divergence as the objective function for on-policy distillation. Specifically, we minimize the divergence between the student's distribution ($\pi_\theta$) and the teacher's ($\pi_{\text{teacher}}$) at each timestep, conditioned on trajectories sampled by the student:

\begin{equation}
    \mathbb{D}_{KL}\left( \pi_\theta \parallel \pi_{\text{teacher}} \right) = \mathbb{E}_{x \sim \pi_{\theta}} \left[ \log \pi_{\theta}(x_{t+1} | x_{1..t}) - \log \pi_{\text{teacher}}(x_{t+1} | x_{1..t}) \right]
\end{equation}

 Inspired by \cite{xu2025kdrl, thinkingmachines2025_onpolicy}, we optimize the policy $\pi_\theta$ directly using policy gradient methods. For each input prompt $q$, we sample a trajectory $o$, and employ the reverse KL divergence between the student policy $\pi_\theta$ and the teacher policy $\pi_{\text{teacher}}$ as the reward signal. The objective is to maximize the expected reward via gradient ascent, with the gradient estimated as follows:
 $$\nabla_\theta J(\theta) = \mathbb{E}_{o \sim \mathcal{D}, x \sim \pi_\theta} \left[ \sum_{t=1}^{T} \nabla_\theta \log \pi_\theta(x_t | o, x_{<t}) \cdot R_{KL}(o, x) \right]$$
where $R_{KL}(o, x)$ denotes the per-token reward derived from the teacher's distribution. To mitigate numerical instability caused by extreme log-probability ratios, we apply a clipping mechanism to the reverse KL divergence. Specifically, the final reward is defined as:
\begin{equation}
    R_{KL}(o, x) = \text{clip}\left( -\mathbb{D}_{KL}(\pi_\theta || \pi_{\text{teacher}}), \alpha, \beta \right)
\end{equation}
where $\alpha$ and $\beta$ represent the lower and upper clipping thresholds, respectively. This constraint prevents outlier reward signals from destabilizing the training process, thereby enabling the model to robustly recover the teacher's behavioral distribution through online trajectory sampling. The comprehensive pipeline is illustrated in Figure~\ref{fig:onpolicy-distillation-pipeline-single}.

\paragraph{On-Policy Distillation on General-Domain.} 

To recover general-domain capabilities, we employ the original Qwen3 model (of the same parameter scale) as a teacher model $\pi_{\text{teacher}}$ to supervise our SFT-refined policy $\pi_\theta$.

A critical challenge lies in the vocabulary discrepancy: the teacher model cannot recognize the itemic tokens in our expanded vocabulary. A naive approach is discarding any trajectory $o$ that contains itemic tokens. However, since the reward is estimated via per-token reverse KL, which is inherently a biased approximation, simply dropping these samples introduces significant sampling bias. As training progresses, the model may gradually increase the probabilities of itemic tokens, eventually leading to training collapse.

To address this, we adopt the following robust distillation strategy:
\begin{itemize}
    \item \textbf{Prompt Selection}: All queries $q$ are sampled exclusively from general-domain datasets. Under these prompts, the policy model is expected to generate pure text without itemic tokens.
    
    \item \textbf{Itemic Token Penalty \& Truncation}: If a sampled trajectory $o$ contains an itemic token at step $t$, we set $\log \pi_{\text{teacher}}(x_t | x_{<t})$ to a minimal value (e.g., $-1e9$) to simulate a zero-probability for that token, and truncate the trajectory after $t$. This penalty, stabilized by the reward clipping mechanism, provides an immediate and strong negative signal for itemic tokens in general-domain contexts.
    
    \item \textbf{Enhanced Exploration}: During on-policy sampling, we employ a relatively high temperature coefficient. This encourages the policy model to explore the vocabulary space (including the itemic tokens), allowing the distillation process to actively identify and correct errors of itemic token activations in general-domain tasks.
\end{itemize}    

Through this distillation process, we sample 200K general-domain questions from the SFT dataset. To better recover the instruction-following capabilities on thinking of the original Qwen3, we follow the methodology described in the Qwen3 technical report \citep{qwen3} and randomly append a suffix (\textbf{\texttt{/think}}, \textbf{\texttt{/no\_think}}, or an empty string) to the user prompt. This strategy is designed to align the model's behavior with the forced-thinking, non-thinking, and auto-thinking paradigms, ensuring robust control over its reasoning processes across diverse scenarios.

After this stage, we effectively bridge the gap in general intelligence while maintaining the model's specialized recommendation performance.

\subsection{Reinforcement Learning for Recommendation}

While on-policy distillation effectively restores the model's general reasoning capabilities, it does not directly optimize the discrete ranking metrics (e.g., Recall or NDCG) that define recommendation quality. Supervised fine-tuning (SFT) primarily focuses on maximizing the likelihood of ground-truth sequences, often suffering from exposure bias and failing to distinguish between "near-misses" and irrelevant recommendations. To bridge this gap, we introduce a final stage of \textbf{Recommendation-oriented Reinforcement Learning (Rec-RL)}. 

\paragraph{Group Relative Policy Optimization}
We employ \textit{Group Relative Policy Optimization} (GRPO)~\cite{shao2024deepseekmath} as our reinforcement learning framework. Unlike traditional Actor-Critic algorithms (e.g., PPO) that require a separate critic model to estimate state values, GRPO computes the advantage of a response relative to a group of sampled trajectories for the same prompt. This significantly reduces computational overhead while maintaining stability. 

Formally, for each recommendation prompt $q$, we sample a group of $G$ candidate responses $\{R_1, R_2, \dots, R_G\}$ from the current policy $\pi_\theta$. The objective is to maximize the following:
\begin{equation}
    \mathcal{L}_{GRPO}(\theta) = \frac{1}{G} \sum_{i=1}^{G} \left( \text{Adv}_i \cdot \log \pi_\theta(R_i | q) \right) - \beta \cdot KL(\pi_\theta || \pi_{ref})
\end{equation}
where $\pi_{ref}$ is the model obtained after on-policy distillation, and $\text{Adv}_i$ is the relative advantage calculated by normalizing the rewards within the group.

\paragraph{Rule-based Recommendation Reward.}
To align the model directly with the ranking accuracy, we design a sparse, rule-based reward function $r(R_i)$ focused on "Hit" events. For the 5 core recommendation tasks—\textit{Short Video Rec}, \textit{Ad Rec}, \textit{Product Rec}, \textit{Interactive Rec}, and \textit{Label-Conditional Rec}—the reward is defined as:
\begin{equation}
    r(R_i) = 
    \begin{cases} 
      +1.0 & \text{if the target itemic token } s \in R_i \\
      0.0 & \text{otherwise}
    \end{cases}
\end{equation}
By sampling multiple candidate sequences (groups) for each user interaction history, GRPO encourages the model to assign higher probability mass to itemic tokens that result in successful hits, effectively performing "Soft Ranking" within the generative space.

\paragraph{Implementation Details.}
We initialize the RL trainer with the model post-distillation. To ensure the model does not sacrifice its restored general intelligence for domain-specific precision, we maintain a strict KL penalty ($\beta$) against $\pi_{ref}$. We utilize the same dataset as the SFT stage, yet observe continuous improvement in recommendation performance throughout RL training. After this stage, we observe a significant boost on recommendation metrics, demonstrating that Rec-RL effectively aligns the language model's generative behavior with the goals of a recommender system.

\section{Evaluation}


In this section, we present our detailed experimental results. We first demonstrate the comprehensive capabilities of OneRec-Foundation on \thebench. Next, we validate the generalization ability of OneRec-Foundation on the Amazon dataset. Finally, we conduct multiple ablation studies to analyze the impact of different components on the model.

\clearpage
\definecolor{lightorange}{rgb}{1, 0.96, 0.92}

\newcommand{\best}[1]{\cellcolor{lightorange}\textbf{#1}}

\begin{table*}[t]
  \setlength{\abovecaptionskip}{0.05cm}
  \setlength{\belowcaptionskip}{0cm}
  \caption{Unified performance comparison across all tasks. For each task, the \textbf{bold}
results highlight the best results (with an orange background), while the second-best ones are \underline{underlined}. Missing entries are marked as ``--''. }
  \label{tab:BigUnifiedTable}

  \footnotesize
  \newcolumntype{C}{>{\centering\arraybackslash}p{0.8cm}}
  \setlength{\tabcolsep}{1.8mm}

  \resizebox{\textwidth}{!}{
  \begin{tabular}{ll|CCCCCCCCCCc}
    \toprule \textbf{Task}
      & \textbf{Metric}
      & \rotatebox{60}{\textbf{SASRec}}
      & \rotatebox{60}{\textbf{BERT4Rec}}
      & \rotatebox{60}{\textbf{GRU4Rec}}
      & \rotatebox{60}{\textbf{HSTU}}
      & \rotatebox{60}{\textbf{ReaRec}}
      & \rotatebox{60}{\textbf{TIGER}}
      & \rotatebox{60}{\textbf{LC-Rec-8B}}
      & \rotatebox{60}{\textbf{OneRec-1.7B}}
      & \rotatebox{60}{\textbf{OneRec-8B}}
      & \rotatebox{60}{\textbf{OneRec-1.7B-Pro}}
      & \rotatebox{60}{\textbf{OneRec-8B-Pro}} \\
    \midrule

    \multirow{3}{*}{\textbf{Short Video Rec}}
      & Pass@1    & 0.0045 & 0.0040 & 0.0051 & 0.0043 & 0.0052 & 0.0168 & 0.0341 & 0.0496 & \underline{0.0542} & 0.0456 & \best{0.0548} \\
      & Pass@32   & 0.1003 & 0.0951 & 0.0993 & 0.1010 & 0.1002 & 0.1061 & 0.1306 & 0.1710 & \underline{0.2104} & 0.1706 & \best{0.2122} \\
      & Recall@32 & 0.0119 & 0.0113 & 0.0117 & 0.0119 & 0.0120 & 0.0132 & 0.0180 & 0.0272 & \underline{0.0355} & 0.0274 & \best{0.0369} \\
    \midrule

    \multirow{3}{*}{\textbf{Ad Rec}}
      & Pass@1    & 0.0044 & 0.0061 & 0.0059 & 0.0076 & 0.0035 & 0.0125 & 0.0197 & 0.0169 & \underline{0.0219} & 0.0190 & \best{0.0259} \\
      & Pass@32   & 0.0980 & 0.1225 & 0.1102 & 0.1266 & 0.1054 & 0.1769 & 0.2096 & 0.2037 & \underline{0.2490} & 0.2126 & \best{0.2700} \\
      & Recall@32 & 0.0293 & 0.0381 & 0.0336 & 0.0409 & 0.0327 & 0.0581 & 0.0723 & 0.0707 & \underline{0.0877} & 0.0735 & \best{0.0964} \\
    \midrule

    \multirow{3}{*}{\textbf{Product Rec}}
      & Pass@1    & 0.0052 & 0.0054 & 0.0047 & 0.0055 & 0.0030 & 0.0120 & 0.0178 & 0.0144 & \underline{0.0187} & 0.0158 & \best{0.0223} \\
      & Pass@32   & 0.0914 & 0.0936 & 0.0821 & 0.0914 & 0.0907 & 0.1276 & 0.1809 & 0.1571 & \underline{0.1971} & 0.1761 & \best{0.2290} \\
      & Recall@32 & 0.0175 & 0.0193 & 0.0161 & 0.0178 & 0.0189 & 0.0283 & 0.0416 & 0.0360 & \underline{0.0470} & 0.0405 & \best{0.0538} \\
    \midrule

    \multirow{3}{*}{\textbf{Label-Cond. Rec}}
      & Pass@1    & 0.0026 & 0.0026 & 0.0032 & 0.0026 & 0.0027 & 0.0044 & 0.0079 & 0.0064 & \underline{0.0097} & 0.0067 & \best{0.0099} \\      
      & Pass@32   & 0.0380 & 0.0372 & 0.0393 & 0.0383 & 0.0381 & 0.0337 & 0.0420 & \underline{0.0431} & 0.0535 & 0.0420 & \best{0.0549} \\
      & Recall@32 & 0.0140 & 0.0135 & 0.0143 & 0.0139 & 0.0137 & 0.0123 & 0.0170 & 0.0184 & \underline{0.0228} & 0.0182 & \best{0.0235} \\
    \midrule
    \multirow{1}{*}{\textbf{Label Pred.}}
      & AUC  & 0.6244 & 0.6598 & 0.6640 & 0.6581 & 0.6204 &  \underline{0.6675} & 0.6139 & 0.6184 & 0.6615 & 0.6071 & \best{0.6912} \\
    \midrule
    \multirow{3}{*}{\textbf{Interactive Rec}}
      & Pass@1    & -- & -- & -- & -- & -- & -- & 0.0890 & 0.0660 & \underline{0.1230} & 0.0800 & \best{0.1250} \\
      & Pass@32   & -- & -- & -- & -- & -- & -- & 0.3730 & 0.3170 & \underline{0.4570} & 0.3370 & \best{0.5080} \\
      & Recall@32 & -- & -- & -- & -- & -- & -- & 0.2394 & 0.1941 & \underline{0.3032} & 0.2024 & \best{0.3458} \\
    \midrule

    \multirow{1}{*}{\textbf{Item Understand.}}
      & LLM-Judge Score & -- & -- & -- & -- & -- & -- & 0.2517 & 0.3175 & \underline{0.3202} & 0.3133 & \best{0.3209} \\

    \midrule

    \multirow{1}{*}{\textbf{Rec. Explanation}}
      & LLM-Judge Score & -- & -- & -- & -- & -- & -- & \underline{3.9350} & 3.3540 & 3.6774 & 3.5060 & \best{4.0381} \\
      
    \bottomrule
  \end{tabular}}
\end{table*}

\begin{table*}[]
  \setlength{\abovecaptionskip}{0.05cm}
  \setlength{\belowcaptionskip}{-0.2cm}
  \caption{Performance comparison on general capability (Thinking). For each task, the \textbf{bold} results highlight the best results, while the second-best ones are \underline{underlined}.}
  \label{tab:GeneralTableThink}

  \footnotesize
  \setlength{\tabcolsep}{1.2mm}

  \resizebox{\textwidth}{!}{
      \begin{tabular}{ll|cccccc}
        \toprule
        \textbf{Category} & \textbf{Task}
        & \textbf{Qwen3-1.7B}
        & \textbf{OneRec-1.7B}
        & \textbf{OneRec-1.7B-Pro}
        & \textbf{Qwen3-8B}
        & \textbf{OneRec-8B}
        & \textbf{OneRec-8B-Pro} \\
        \midrule

        \multirow{3}{*}{\textbf{\makecell[l]{Math \& Text\\Reasoning}}}
        & MATH-500 & 0.8780 & 0.8840 & 0.8840 & \textbf{0.9520} & \underline{0.9460} & 0.9380 \\
        & GSM8K & 0.9121 & 0.8984 & 0.8999 & \underline{0.9568} & \textbf{0.9575} & \textbf{0.9575} \\
        & AIME'24 & 0.4938 & 0.4104 & 0.4146 & \textbf{0.7917} & \underline{0.7250} & 0.7188 \\
        \midrule

        \multirow{2}{*}{\textbf{General Tasks}}
        & MMLU-Pro & 0.5422 & 0.3548 & 0.3932 & \textbf{0.7235} & \underline{0.5342} & 0.5204 \\
        & GPQA-Diamond & 0.3788 & 0.3232 & 0.3333 & \textbf{0.5606} & 0.5000 & \underline{0.5051} \\
        \midrule

        \textbf{Alignment Tasks}
        & IFEVAL$_{\text{strict prompt}}$ & 0.6969 & 0.5471 & 0.5416 & \textbf{0.8577} & \underline{0.7893} & 0.7634 \\
        \midrule

        \textbf{Coding}
        & LiveCodeBench v5 & 0.3907 & 0.2832 & 0.2832 & \textbf{0.5484} & \underline{0.4910} & 0.4667 \\

        \bottomrule
    \end{tabular}
  }
\end{table*}

\begin{table*}[]
  \setlength{\abovecaptionskip}{0.2cm}
  \setlength{\belowcaptionskip}{0cm}
  \caption{Performance comparison on general capability (Non-Thinking). For each task, the \textbf{bold} results highlight the best results, while the second-best ones are \underline{underlined}.}
  \label{tab:GeneralTableNoThink}

  \footnotesize
  \setlength{\tabcolsep}{1.2mm}

  \resizebox{\textwidth}{!}{
      \begin{tabular}{ll|cccccc}
        \toprule
        \textbf{Category} & \textbf{Task}
        & \textbf{Qwen3-1.7B}
        & \textbf{OneRec-1.7B}
        & \textbf{OneRec-1.7B-Pro}
        & \textbf{Qwen3-8B}
        & \textbf{OneRec-8B}
        & \textbf{OneRec-8B-Pro} \\
        \midrule

        \multirow{3}{*}{\textbf{\makecell[l]{Math \& Text\\Reasoning}}}
        & MATH-500 & 0.6980 & 0.7060 & 0.6940 & \textbf{0.8380} & \underline{0.8240} & 0.7980 \\
        & GSM8K & 0.8218 & 0.8036 & 0.8158 & \underline{0.9303} & \textbf{0.9310} & 0.9196 \\
        & AIME'24 & 0.1313 & 0.1271 & 0.1250 & \textbf{0.2729} & \underline{0.2417} & 0.2271 \\
        \midrule

        \multirow{2}{*}{\textbf{General Tasks}}
        & MMLU-Pro & 0.4384 & 0.3072 & 0.2804 & \textbf{0.6632} & \underline{0.5795} & 0.4521 \\
        & GPQA-Diamond & 0.3030 & 0.3131 & 0.2778 & \underline{0.3990} & \textbf{0.4040} & 0.3939 \\
        \midrule

        \textbf{Alignment Tasks}
        & IFEVAL$_{\text{strict prompt}}$ & 0.6747 & 0.4769 & 0.5250 & \textbf{0.8392} & \underline{0.7357} & 0.7098 \\
        \midrule

        \textbf{Coding}
        & LiveCodeBench v5 & 0.1219 & 0.1219 & 0.1147 & \textbf{0.2760} & \underline{0.2401} & \underline{0.2401} \\

        \bottomrule
    \end{tabular}
  }
\end{table*}

\clearpage
\subsection{Experimental Settings}



\paragraph{RecIF-Bench.}
To demonstrate the effectiveness of OneRec-Foundation, we compare it against two groups of competitive baselines: (1) \textit{Discriminative recommender models}, including BERT4Rec~\citep{sun2019bert4rec}, GRU4Rec~\citep{hidasi2016sessionbasedrecommendationsrecurrentneural}, SASRec~\citep{kang2018self}, HSTU~\citep{zhai2024actions}, and ReaRec~\citep{tang2025think}; and (2) \textit{Generative recommender models}, such as TIGER~\citep{rajput2023recommender} and LC-Rec~\citep{zheng2024adapting}. We adapted these baseline methods with task-specific modifications to support different RecIF tasks. Notably, these discriminative-based approaches are inherently task-specific, with each task requiring a separately trained model. To ensure a fair comparison at the foundation model scale, we implement LC-Rec as ``LC-Rec-8B'' using a comparable Qwen3-8B. We evaluate all methods using the metrics in Section~\ref{subsec:evaluation_metrics}, with comprehensive results reported in Table~\ref{tab:BigUnifiedTable}.

\paragraph{Amazon.}
To evaluate the generalization capability of our model, we utilize ten real-world datasets from the popular Amazon review benchmark~\citep{mcauley2015image}, covering diverse domains: \textit{Baby}, \textit{Beauty}, \textit{Cell Phones and Accessories}, \textit{Grocery and Gourmet Food}, \textit{Health and Personal Care}, \textit{Home and Kitchen}, \textit{Pet Supplies}, \textit{Sports and Outdoors}, \textit{Tools and Home Improvement}, and \textit{Toys and Games}.
We adopt the same baselines as mentioned above.
Consistent with previous works~\citep{rajput2023recommender, wang2024learnableitemtokenizationgenerative}, we focus on the sequential recommendation setting. For data pre-processing, we discard sparse users and items with fewer than 5 interactions. We employ the \textit{leave-one-out} strategy~\citep{rajput2023recommender, wang2024learnableitemtokenizationgenerative} to split the datasets for training and evaluation. For performance comparison, we report Recall@$K$ and NDCG@$K$ with $K \in \{5, 10\}$.


\subsection{Main Results on \thebench}












Based on the evaluation results, we highlight some key conclusions on \thebench:
\begin{itemize}
    \item \textbf{State-of-the-Art Recommendation Performance}: OneRec-Foundation consistently outperforms all baselines across the vast majority of tasks. Notably, the results demonstrate scaling from two dimensions: (1) \textit{Data scaling}: OneRec-Pro consistently surpasses OneRec at the same model size; (2) \textit{Model scaling}: 8B models outperform 1.7B models across all variants.
    \item \textbf{Trade-off on General Capabilities}: As shown in Table~\ref{tab:GeneralTableThink} and Table~\ref{tab:GeneralTableNoThink}, our model successfully retains most of the general capabilities of the Qwen3 backbone, especially with minimal degradation on mathematical benchmarks. Nevertheless, we observe a performance trade-off in general knowledge and recommendation capabilities. This suggests that while the distillation process effectively preserves reasoning proficiency, the limited diversity of the general data used may have constrained the model's broader capabilities, indicating that a more refined data strategy is required to better balance recommendation and general capabilities.
\end{itemize}

\subsection{Transfer Learning on Amazon Benchmark}
To evaluate the transferability of OneRec-Foundation, we conduct comprehensive experiments on the Amazon benchmark, encompassing 10 distinct domains. These experiments serve to rigorously validate whether our foundation model, pre-trained on diverse open-domain data, confers a fundamental transfer advantage for modeling specific downstream recommendation distributions.
\definecolor{lightorange}{rgb}{1, 0.96, 0.92} 

\setlength{\fboxsep}{1pt} 


\begin{table}[t]

  \centering

  \caption{Cross-Domain generalization performance on Amazon domains. Best results are indicated with bold fonts and orange background.}

  \label{tab:amazon_domains_performance}

  \setlength{\tabcolsep}{3mm}

  \resizebox{\textwidth}{!}{%

  \begin{tabular}{c|c|cccccccccc}

    \toprule

    \textbf{Model} & \textbf{Metric}

    & \textbf{Baby}

    & \textbf{Beauty}

    & \textbf{Cell}

    & \textbf{Grocery}

    & \textbf{Health}

    & \textbf{Home}

    & \textbf{Pet}

    & \textbf{Sports}

    & \textbf{Tools}

    & \textbf{Toys} \\

    \midrule

\multirow{4}{*}{\textbf{SASRec}} & R@5 & \underline{0.0232} & 0.0393 & 0.0482 & 0.0480 & 0.0295 & 0.0133 & 0.0377 & 0.0240 & 0.0269 & 0.0420 \\

 & R@10 & \underline{0.0381} & 0.0639 & 0.0782 & 0.0789 & 0.0506 & 0.0212 & 0.0607 & 0.0389 & 0.0437 & 0.0658 \\

 & N@5 & 0.0137 & 0.0209 & 0.0281 & 0.0262 & 0.0173 & 0.0070 & 0.0222 & 0.0130 & 0.0149 & 0.0217 \\

 & N@10 & 0.0185 & 0.0289 & 0.0378 & 0.0361 & 0.0242 & 0.0098 & 0.0296 & 0.0178 & 0.0203 & 0.0294 \\

\midrule

\multirow{4}{*}{\textbf{BERT4Rec}} & R@5 & 0.0117 & 0.0219 & 0.0325 & 0.0307 & 0.0204 & 0.0063 & 0.0218 & 0.0151 & 0.0145 & 0.0200 \\

 & R@10 & 0.0228 & 0.0419 & 0.0569 & 0.0534 & 0.0353 & 0.0113 & 0.0412 & 0.0261 & 0.0264 & 0.0362 \\

 & N@5 & 0.0065 & 0.0120 & 0.0190 & 0.0174 & 0.0117 & 0.0038 & 0.0123 & 0.0083 & 0.0083 & 0.0102 \\

 & N@10 & 0.0101 & 0.0185 & 0.0268 & 0.0247 & 0.0165 & 0.0054 & 0.0186 & 0.0119 & 0.0121 & 0.0154 \\

\midrule

\multirow{4}{*}{\textbf{GRU4Rec}} & R@5 & 0.0202 & 0.0322 & 0.0430 & 0.0362 & 0.0256 & 0.0090 & 0.0264 & 0.0174 & 0.0176 & 0.0266 \\

 & R@10 & 0.0346 & 0.0539 & 0.0676 & 0.0591 & 0.0423 & 0.0156 & 0.0449 & 0.0278 & 0.0305 & 0.0453 \\

 & N@5 & 0.0124 & 0.0201 & 0.0275 & 0.0230 & 0.0164 & 0.0058 & 0.0163 & 0.0110 & 0.0116 & 0.0171 \\

 & N@10 & 0.0170 & 0.0271 & 0.0355 & 0.0303 & 0.0217 & 0.0079 & 0.0222 & 0.0144 & 0.0158 & 0.0231 \\

\midrule

\multirow{4}{*}{\textbf{HSTU}} & R@5 & 0.0226 & 0.0456 & 0.0475 & 0.0458 & 0.0330 & 0.0134 & 0.0362 & 0.0227 & 0.0231 & 0.0489 \\

 & R@10 & 0.0350 & 0.0643 & 0.0725 & 0.0712 & 0.0485 & 0.0197 & 0.0521 & 0.0347 & 0.0337 & 0.0649 \\

 & N@5 & \underline{0.0156} & 0.0308 & 0.0314 & 0.0297 & 0.0215 & 0.0092 & 0.0239 & 0.0151 & 0.0159 & 0.0339 \\

 & N@10 & \underline{0.0196} & 0.0368 & 0.0395 & 0.0378 & 0.0265 & 0.0112 & 0.0290 & 0.0190 & 0.0193 & 0.0391 \\

\midrule

\multirow{4}{*}{\textbf{ReaRec}} & R@5 & 0.0197 & 0.0488 & 0.0444 & 0.0454 & 0.0326 & 0.0150 & 0.0299 & 0.0231 & 0.0219 & \underline{0.0517} \\

 & R@10 & 0.0320 & 0.0702 & 0.0711 & 0.0730 & 0.0481 & 0.0210 & 0.0486 & 0.0348 & 0.0310 & \underline{0.0706} \\

 & N@5 & 0.0123 & \underline{0.0341} & 0.0269 & 0.0289 & 0.0213 & 0.0101 & 0.0189 & 0.0152 & 0.0143 & \underline{0.0369} \\

 & N@10 & 0.0163 & 0.0409 & 0.0355 & 0.0378 & 0.0263 & 0.0121 & 0.0249 & 0.0189 & 0.0173 & \underline{0.0430} \\

\midrule

\multirow{4}{*}{\textbf{TIGER}} & R@5 & 0.0191 & 0.0413 & 0.0540 & 0.0447 & 0.0328 & 0.0142 & 0.0343 & 0.0216 & 0.0228 & 0.0367 \\

 & R@10 & 0.0318 & 0.0628 & 0.0786 & 0.0691 & 0.0534 & 0.0216 & 0.0542 & 0.0331 & 0.0344 & 0.0527 \\

 & N@5 & 0.0125 & 0.0277 & 0.0350 & 0.0295 & 0.0222 & 0.0094 & 0.0232 & 0.0145 & 0.0148 & 0.0255 \\

 & N@10 & 0.0162 & 0.0346 & 0.0429 & 0.0373 & 0.0289 & 0.0118 & 0.0295 & 0.0182 & 0.0184 & 0.0307 \\

\midrule

\multirow{4}{*}{\textbf{LC-Rec}} & R@5 & 0.0232 & \underline{0.0495} & \underline{0.0585} & \underline{0.0501} & \underline{0.0412} & \underline{0.0199} & \underline{0.0388} & \underline{0.0269} & \underline{0.0288} & 0.0350 \\

 & R@10 & 0.0344 & \underline{0.0764} & \underline{0.0883} & \underline{0.0790} & \underline{0.0616} & \underline{0.0293} & \underline{0.0612} & \underline{0.0418} & \underline{0.0438} & 0.0549 \\

 & N@5 & 0.0151 & 0.0338 & \underline{0.0392} & \underline{0.0328} & \underline{0.0272} & \underline{0.0138} & \underline{0.0247} & \underline{0.0177} & \underline{0.0187} & 0.0221 \\

 & N@10 & 0.0187 & \underline{0.0424} & \underline{0.0488} & \underline{0.0421} & \underline{0.0338} & \underline{0.0168} & \underline{0.0320} & \underline{0.0225} & \underline{0.0235} & 0.0285 \\

\midrule

\multirow{4}{*}{\textbf{Ours}} & R@5 & 
\cellcolor{lightorange}\textbf{0.0352} & 
\cellcolor{lightorange}\textbf{0.0646} & \cellcolor{lightorange}\textbf{0.0717} & \cellcolor{lightorange}\textbf{0.0688} & \cellcolor{lightorange}\textbf{0.0534} & \cellcolor{lightorange}\textbf{0.0279} & \cellcolor{lightorange}\textbf{0.0563} & \cellcolor{lightorange}\textbf{0.0365} & \cellcolor{lightorange}\textbf{0.0412} & \cellcolor{lightorange}\textbf{0.0693} \\

 & R@10 & \cellcolor{lightorange}\textbf{0.0513} & \cellcolor{lightorange}\textbf{0.0924} & \cellcolor{lightorange}\textbf{0.1036} & \cellcolor{lightorange}\textbf{0.1029} & \cellcolor{lightorange}\textbf{0.0768} & \cellcolor{lightorange}\textbf{0.0390} & \cellcolor{lightorange}\textbf{0.0834} & \cellcolor{lightorange}\textbf{0.0547} & \cellcolor{lightorange}\textbf{0.0593} & \cellcolor{lightorange}\textbf{0.0953} \\

 & N@5 & \cellcolor{lightorange}\textbf{0.0238} & \cellcolor{lightorange}\textbf{0.0456} & \cellcolor{lightorange}\textbf{0.0490} & \cellcolor{lightorange}\textbf{0.0460} & \cellcolor{lightorange}\textbf{0.0376} & \cellcolor{lightorange}\textbf{0.0202} & \cellcolor{lightorange}\textbf{0.0389} & \cellcolor{lightorange}\textbf{0.0252} & \cellcolor{lightorange}\textbf{0.0295} & \cellcolor{lightorange}\textbf{0.0496} \\

 & N@10 & \cellcolor{lightorange}\textbf{0.0289} & \cellcolor{lightorange}\textbf{0.0545} & \cellcolor{lightorange}\textbf{0.0593} & \cellcolor{lightorange}\textbf{0.0570} & \cellcolor{lightorange}\textbf{0.0452} & \cellcolor{lightorange}\textbf{0.0237} & \cellcolor{lightorange}\textbf{0.0476} & \cellcolor{lightorange}\textbf{0.0310} & \cellcolor{lightorange}\textbf{0.0354} & \cellcolor{lightorange}\textbf{0.0579} \\

\midrule

\midrule

\multicolumn{2}{c|}{\textbf{Improve (\%) R@10}} 

& 34.6\,\raisebox{0.3ex}{\scalebox{0.7}{$\uparrow$}} 

& 20.9\,\raisebox{0.3ex}{\scalebox{0.7}{$\uparrow$}} 

& 17.3\,\raisebox{0.3ex}{\scalebox{0.7}{$\uparrow$}} 

& 30.3\,\raisebox{0.3ex}{\scalebox{0.7}{$\uparrow$}} 

& 24.7\,\raisebox{0.3ex}{\scalebox{0.7}{$\uparrow$}} 

& 33.1\,\raisebox{0.3ex}{\scalebox{0.7}{$\uparrow$}} 

& 36.3\,\raisebox{0.3ex}{\scalebox{0.7}{$\uparrow$}} 

& 30.9\,\raisebox{0.3ex}{\scalebox{0.7}{$\uparrow$}} 

& 35.4\,\raisebox{0.3ex}{\scalebox{0.7}{$\uparrow$}} 

& 35.0\,\raisebox{0.3ex}{\scalebox{0.7}{$\uparrow$}} \\

    \bottomrule

  \end{tabular}}

\end{table}

\paragraph{Main Results.}
As detailed in Table~\ref{tab:amazon_domains_performance}, OneRec-Foundation establishes new state-of-the-art results across all 10 datasets. Specifically, our model achieves an average improvement of \textbf{26.8\%} in Recall@10 over the second-best baseline on each domain. These results empirically confirm that large-scale generative pre-training endows the model with robust transfer capabilities that far exceed traditional collaborative filtering approaches.

Through our transfer learning experiments, we identify two critical factors that significantly influence performance: \textbf{Comprehensive Utilization of Pre-trained Knowledge}, where we design adaptive strategies to maximally leverage the collaborative filtering signals and semantic understanding capabilities encoded during pre-training (detailed in Table~\ref{tab:word_sid_word_performance}), and \textbf{Multi-Domain Joint Training}, which enables the model to extract universal recommendation patterns across heterogeneous domains, as demonstrated in Figure~\ref{fig:amazon_comparison}. We elaborate on these two critical aspects in the following subsections.

\subsubsection{Adaptive Strategies for Pre-trained Model Utilization}

A primary challenge in transfer learning is the distributional shift of item identifiers. Our pre-trained tokenizer is optimized on a broad, open-domain corpus, resulting in a codebook that may not granularly distinguish items within a specific vertical like Amazon products. Direct application leads to a high collision rate (>30\%), causing catastrophic information loss. To address this, we systematically explore three strategies to adapt pre-trained representations to the target domain:

\paragraph{Strategy 1: Extended Residual Quantization.}
We extend the hierarchical depth by computing residuals from the pre-trained third layer and applying Finite Scalar Quantization (FSQ)~\citep{mentzer2023finite} to generate a fourth-layer code, reducing collisions to 3.05\%. Remaining collisions are resolved via popularity-based decoding. This strategy achieves a 10.0\% improvement in average R@10 over LC-Rec, validating effective transfer of collaborative filtering knowledge. However, the non-pretrained fourth layer disrupts the original hierarchical semantics, motivating alternative approaches.

\paragraph{Strategy 2: Text-Only Adaptation.}
We bypass itemic tokens entirely, representing each item via 5 distinctive keywords extracted from its metadata~\citep{zhang2026unleashing}, reducing collisions to 4.27\%. This strategy achieves an 18.8\% improvement in average R@10 over Extended Residual Quantization: the model's linguistic core remains intact, enabling robust semantic understanding, while natural language representations prove more expressive in narrow domains. However, this approach sacrifices collaborative filtering signals embedded in pre-trained itemic tokens.

\paragraph{Strategy 3: Text-Augmented Itemic Tokens.}
We concatenate the original three-layer pre-trained itemic tokens with keyword representations: \texttt{[itemic\_tokens] + [keywords]}. Critically, we preserve the original pre-trained itemic tokens without structural extension, maintaining hierarchical semantics. Keywords provide semantic disambiguation (collision rate 0.47\%) and enable full utilization of linguistic capabilities. Table~\ref{tab:word_sid_word_performance} shows that this strategy achieves state-of-the-art performance across nearly all datasets. The consistent gains validate that effective transfer learning requires maximizing utilization of the foundation model's diverse capabilities—collaborative filtering, knowledge, and semantic understanding—while strictly preserving pre-trained structural integrity.

\begin{table}[htbp]
  \centering
  \caption{Performance Comparison of Adaptive Strategies for Pre-trained Model Utilization. Extended Residual Quantization (collision rate: 3.05\%), Text-Only Adaptation (collision rate: 4.27\%), and Text-Augmented Itemic Tokens (collision rate: 0.47\%).}
  \label{tab:word_sid_word_performance}
  \setlength{\tabcolsep}{2mm}

  \resizebox{\textwidth}{!}{%
  \begin{tabular}{c|c|cccccccccc}
    \toprule
    \textbf{Strategy} & \textbf{Metric}
    & \textbf{\makecell{Baby}}
    & \textbf{\makecell{Beauty}}
    & \textbf{\makecell{Cell}}
    & \textbf{\makecell{Grocery}}
    & \textbf{\makecell{Health}}
    & \textbf{\makecell{Home}}
    & \textbf{\makecell{Pet}}
    & \textbf{\makecell{Sports}}
    & \textbf{\makecell{Tools}}
    & \textbf{\makecell{Toys}} \\
    \midrule

    \multirow{4}{*}{\textbf{\makecell{Extended\\Residual\\Quantization}}}
      & R@5  & 0.0288 & 0.0534 & 0.0574 & 0.0562 & 0.0479 & 0.0227 & 0.0518 & 0.0315 & 0.0350 & 0.0511 \\
      & R@10 & 0.0407 & 0.0799 & 0.0830 & 0.0861 & 0.0673 & 0.0313 & 0.0758 & 0.0447 & 0.0495 & 0.0701 \\
      & N@5  & 0.0201 & 0.0364 & 0.0389 & 0.0383 & 0.0333 & 0.0162 & 0.0356 & 0.0215 & 0.0243 & 0.0360 \\
      & N@10 & 0.0239 & 0.0449 & 0.0471 & 0.0480 & 0.0396 & 0.0190 & 0.0433 & 0.0258 & 0.0289 & 0.0421 \\
    \midrule

    \multirow{4}{*}{\textbf{\makecell{Text-Only\\Adaptation}}}
      & R@5  & 0.0317 & 0.0630 & 0.0688 & 0.0687 & 0.0529 & 0.0285 & 0.0548 & 0.0368 & 0.0414 & 0.0668 \\
      & R@10 & 0.0448 & 0.0883 & 0.0985 & \textbf{0.1048} & 0.0752 & 0.0398 & \textbf{0.0850} & \textbf{0.0548} & \textbf{0.0615} & 0.0931 \\
      & N@5  & 0.0227 & 0.0445 & 0.0473 & 0.0460 & 0.0368 & 0.0199 & 0.0382 & 0.0256 & 0.0288 & 0.0483 \\
      & N@10 & 0.0269 & 0.0526 & 0.0569 & 0.0576 & 0.0440 & 0.0235 & 0.0478 & 0.0314 & 0.0354 & 0.0568 \\
    \midrule

    \multirow{4}{*}{\textbf{\makecell{Text-Augmented\\Itemic Tokens}}}
      & R@5  & \textbf{0.0352} & \textbf{0.0646} & \textbf{0.0717} & \textbf{0.0688} & \textbf{0.0534} & \textbf{0.0285} & \textbf{0.0563} & \textbf{0.0368} & \textbf{0.0414} & \textbf{0.0693} \\
      & R@10 & \textbf{0.0513} & \textbf{0.0924} & \textbf{0.1036} & 0.1029 & \textbf{0.0768} & \textbf{0.0398} & 0.0834 & 0.0547 & 0.0593 & \textbf{0.0953} \\
      & N@5  & \textbf{0.0238} & \textbf{0.0456} & \textbf{0.0490} & \textbf{0.0460} & \textbf{0.0376} & \textbf{0.0202} & \textbf{0.0389} & \textbf{0.0256} & \textbf{0.0295} & \textbf{0.0496} \\
      & N@10 & \textbf{0.0289} & \textbf{0.0545} & \textbf{0.0593} & \textbf{0.0576} & \textbf{0.0452} & \textbf{0.0237} & \textbf{0.0478} & \textbf{0.0314} & \textbf{0.0354} & \textbf{0.0579} \\
    \bottomrule
  \end{tabular}}
\end{table}

\subsubsection{Domain-Specific Training vs. Multi-Domain Joint Training }

Beyond item representation, another critical factor is the training strategy across domains. We compare \textbf{Domain-Specific Training} versus \textbf{Multi-Domain Joint Training} to investigate whether our pre-trained foundation model can benefit from multi-domain knowledge integration, in contrast to traditional generative recommenders like TIGER.

\begin{figure}[t]
    \centering
    \includegraphics[width=\textwidth]{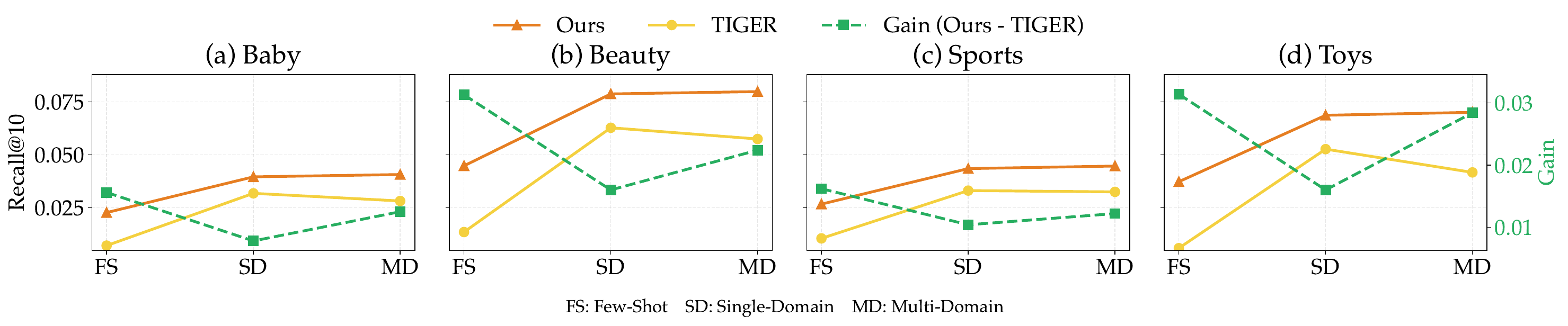}
    \caption{\textbf{Impact of Training Strategies (Domain-Specific vs Multi-Domain Joint) and Few-Shot Learning on Transfer Performance.} We compare OneRec-Foundation (Ours) against TIGER across four Amazon domains under three settings: \textbf{(1)} Few-shot learning with 10\% training data, \textbf{(2)} Full-data training with domain-specific strategy, and \textbf{(3)} Full-data training with joint multi-domain strategy. The green dashed line represents the performance gain (Recall@10 difference) of Ours over TIGER.}
    \label{fig:amazon_comparison}
\end{figure}

The results in Figure~\ref{fig:amazon_comparison} reveal a striking divergence. TIGER exhibits a consistent performance decline under joint training, with an average 10.6\% drop in Recall@10. In stark contrast, OneRec-Foundation demonstrates an average 2.3\% improvement.

This divergence highlights the fundamental advantage of our pre-trained foundation model. Traditional models like TIGER primarily memorize domain-specific collaborative statistics. When confronted with heterogeneous multi-domain data, they struggle to reconcile conflicting patterns, leading to performance degradation.

In contrast, OneRec-Foundation's success stems from the unique combination of rich recommendation knowledge and semantic understanding capabilities acquired during pre-training. This enables the model to extract generalizable patterns rather than memorize domain-specific statistics. Multi-domain joint training further enriches the model by exposing it to diverse interaction patterns, enabling effective cross-domain knowledge transfer. The massive parameter capacity provides sufficient representational space to encode domain-specific nuances while maintaining shared high-level patterns.

\paragraph{Few-Shot Learning: Amplified Transfer Advantage.}
Beyond training strategies, Figure~\ref{fig:amazon_comparison} also reveals that the transfer learning advantage of foundation models becomes \textit{significantly more pronounced} under data scarcity. While OneRec-Foundation surpasses TIGER by an average of \textbf{77.7\%} in Recall@10 with full training data, this gap widens dramatically to \textbf{219.7\%} in the 10\% few-shot regime. Crucially, OneRec-Foundation preserves \textbf{45.2\%} of its full-data performance when restricted to 10\% data, whereas TIGER retains only \textbf{23.0\%}. This striking resilience validates that large-scale pre-training confers robust, transferable representations that enable effective domain adaptation under severe data constraints.

\subsection{Ablation Study}

\subsubsection{Ablation Study on Pre-training Strategies}

\begin{table}[htbp]
  \caption{\textbf{Ablation study on pre-training strategies.} We compare our model with ablated variants. w/o Align: removing the itemic-text alignment stage. Results are reported across three model sizes (0.6B, 1.7B, 8B). Bold indicates the best result for each task within each model size.}
  \label{tab:pretrain-abliation}
  \centering
  \footnotesize
  \setlength{\tabcolsep}{2pt}
  \resizebox{0.75\textwidth}{!}{
  \begin{tabular}{l|l|cc|cc|cc}
    \toprule
    \multirow{2}{*}{\textbf{Task}} & \multirow{2}{*}{\textbf{Metric}} & \multicolumn{2}{c|}{\textbf{0.6B}} & \multicolumn{2}{c|}{\textbf{1.7B}} & \multicolumn{2}{c}{\textbf{8B}} \\
    \cmidrule(lr){3-4} \cmidrule(lr){5-6} \cmidrule(lr){7-8}
    & & Ours & w/o Align & Ours & w/o Align & Ours & w/o Align \\
    \midrule
    \multirow{2}{*}{\textbf{Short Video Rec}} 
    & Pass@32 & \textbf{0.1401} & 0.1397 & \textbf{0.1636} & 0.1605 & \textbf{0.2034} & 0.1933 \\
    & Recall@32 & \textbf{0.0210} & \textbf{0.0210} & \textbf{0.0254} & 0.0251 & \textbf{0.0334} & 0.0310 \\
    \midrule
    \multirow{2}{*}{\textbf{Ad Rec}}
    & Pass@32 & \textbf{0.1740} & 0.1680 & \textbf{0.1961} & 0.1922 & 0.2350 & \textbf{0.2401} \\
    & Recall@32 & \textbf{0.0586} & 0.0569 & \textbf{0.0673} & 0.0669 & 0.0821 & \textbf{0.0841} \\
    \midrule
    \multirow{2}{*}{\textbf{Product Rec}} 
    & Pass@32 & \textbf{0.1139} & 0.1064 & \textbf{0.1512} & 0.1395 & 0.1893 & \textbf{0.1911} \\
    & Recall@32 & \textbf{0.0257} & 0.0243 & \textbf{0.0343} & 0.0312 & \textbf{0.0447} & 0.0442 \\
    \midrule
    \multirow{2}{*}{\textbf{Label-Cond. Rec}} 
    & Pass@32 & \textbf{0.0350} & 0.0343 & \textbf{0.0426} & 0.0401 & \textbf{0.0537} & \textbf{0.0537} \\
    & Recall@32 & \textbf{0.0146} & 0.0145 & \textbf{0.0181} & 0.0171 & 0.0227 & \textbf{0.0230} \\
    \midrule
    \multirow{2}{*}{\textbf{Interactive Rec}} 
    & Pass@32 & \textbf{0.2460} & 0.2360 & \textbf{0.3110} & 0.3050 & \textbf{0.4650} & 0.4490 \\
    & Recall@32 & \textbf{0.1402} & 0.1357 & \textbf{0.1908} & 0.1770 & \textbf{0.3039} & 0.2910 \\
    \midrule
    \textbf{Label Pred.} & AUC & \textbf{0.6488} & 0.5807 & \textbf{0.6392} & 0.5796 & \textbf{0.6879} & 0.6285 \\
    \midrule
    \textbf{Item Understanding}  & LLM-Judge Score & \textbf{0.3174} & 0.3112 & 0.3170 & \textbf{0.3181} & \textbf{0.3225} & 0.3103 \\
    \midrule
    \textbf{Rec. Explanation} & LLM-Judge Score & \textbf{2.9960} & 2.8635 & 3.0922 & \textbf{3.3160} & \textbf{3.9420} & 3.9329 \\
    \bottomrule
  \end{tabular}
}
\end{table}

We conducted an ablation study to quantify the contribution of the \textbf{itemic-text alignment (Stage 1)} within our pre-training pipeline. We compare the full-stage model against an ablated variant (w/o Align) that bypasses this initial phase. Both versions utilized an identical pre-training recipe, including the optimizer, learning-rate schedule, and context length. Since the raw pre-trained checkpoints lack inherent instruction-following capabilities, we applied the Multi-task SFT described in Section \ref{subsec:multi-task-sft} to both variants before benchmarking across three model scales (0.6B, 1.7B, and 8B).

The results, summarized in Table \ref{tab:pretrain-abliation}, reveal that Stage 1 serves as a fundamental semantic bridge for cold-started itemic token embeddings. By aligning these initialized parameters with the pre-trained latent space before full-parameter fine-tuning, Stage 1 establishes a robust semantic foundation, which is particularly necessary for smaller models(0.6B, 1.7B). The marginal gains from Stage 1 scale inversely with model size, likely because larger backbones possess greater inherent generalization. However, this stage remains essential for domain-specific precision, particularly in label prediction and interactive recommendation. These findings underscore that explicit alignment is a prerequisite for optimizing recommendation performance across all model scales.

\subsubsection{Evolution of Model Capabilities Across Post-training Stages}

We also analyze the performance of our model after each key post-training phase: \textit{Multi-task Supervised Fine-tuning}, \textit{On-policy Distillation for General Capability}, and \textit{Reinforcement Learning for Recommendation}. As shown in Table~\ref{tab:gen-post-stage-think}, Table~\ref{tab:gen-post-stage-nothink}, and Table~\ref{tab:reco-posttrain-abliation}, each stage plays a distinct role in balancing the performance between recommendation-domain and general-domain.


\begin{table}[h]
\centering
\caption{\textbf{ Performance comparison on general capability across post-training stages (Thinking).}}
\label{tab:gen-post-stage-think}
\resizebox{\textwidth}{!}{
\begin{tabular}{l|ccccccc}
\toprule
 & \textbf{Math 500} & \textbf{GSM8K} & \textbf{AIME'24} & \textbf{MMLU Pro} & \textbf{GPQA Diamond} & \textbf{IFEVAL} & \textbf{LiveCodeBench} \\
\midrule
\textbf{Qwen3-8B (Base)} & \textbf{0.952} & 0.9568 & \textbf{0.7917} & \textbf{0.7235} & \textbf{0.5606} & \textbf{0.8577} & \textbf{0.5484} \\
\midrule
\textbf{Stage 1: Multi-task SFT}          & 0.936          & 0.9083 & 0.5104          & 0.5307          & 0.4949          & 0.6174          & 0.4516          \\
\textbf{Stage 2: On-Policy Distillation}          & 0.948          & 0.9538 & 0.7125          & 0.5454          & 0.5             & 0.7653          & 0.4659          \\
\textbf{Stage 3: Reinforcement Learning}          & 0.938          & \textbf{0.9575} & 0.7188          & 0.5204          & 0.5051          & 0.7634          & 0.4667          \\
\bottomrule
\end{tabular}
}
\end{table}

\begin{table}[h]
\centering
\caption{\textbf{ Performance comparison on general capability across post-training stages (Non-Thinking).}}
\label{tab:gen-post-stage-nothink}
\resizebox{\textwidth}{!}{
\begin{tabular}{l|ccccccc}
\toprule
 & \textbf{Math 500} & \textbf{GSM8K} & \textbf{AIME'24} & \textbf{MMLU Pro} & \textbf{GPQA Diamond} & \textbf{IFEVAL} & \textbf{LiveCodeBench} \\
\midrule
\textbf{Qwen3-8B} & 0.838 & \textbf{0.9303} & \textbf{0.2729} & \textbf{0.6632} & 0.399 & \textbf{0.8392} & 0.276 \\
\midrule
\textbf{Stage 1: Multi-task SFT}          & \textbf{0.876} & 0.906 & 0.0688          & 0.4909          & 0.3384          & 0.5638          & 0.1756          \\
\textbf{Stage 2: On-Policy Distillation}  & 0.848          & 0.9234 & 0.2521          & 0.583           & \textbf{0.4091} & 0.7689          & \textbf{0.2545} \\
\textbf{Stage 3: Reinforcement Learning}  & 0.798          & 0.9196 & 0.2271          & 0.4521          & 0.3939          & 0.7098          & 0.2401          \\
\bottomrule
\end{tabular}
}
\end{table}

\textbf{Impact of On-policy Distillation on General Capabilities.} As shown in Table~\ref{tab:gen-post-stage-think} and Table~\ref{tab:gen-post-stage-nothink}, a comparison between Stage 1 (Multi-task SFT) and Stage 2 (On-policy Distillation) reveals that on-policy distillation significantly restores general capabilities, effectively realigning the model with the Qwen3 baseline on most general benchmarks. Despite this marked improvement, a performance gap persists relative to the original Qwen3 base model across several metrics. This gap is likely attributable to the current data composition and quality during the distillation phase, suggesting that further refinement of the data strategy is required to fully bridge the remaining gap.

Interestingly, after the Multi-task SFT stage, we observed that several metrics in the "Non-Thinking" mode are unexpectedly high, occasionally even surpassing the Qwen3 base model. Qualitative results indicate this is due to instruction drift, where the model disregards the \textbf{\texttt{/no\_think}} tag and generates "thinking" trajectories (CoT), leading to inflated scores. This issue is effectively mitigated through On-policy Distillation, which restores the model’s ability to faithfully switch between distinct reasoning modes.

\textbf{Advancements through RL for Recommendation.}
The final reinforcement learning stage demonstrates targeted improvements on core recommendation tasks. As illustrated in Table~\ref{tab:reco-posttrain-abliation}, the RL-trained model achieves consistent gains on Reco. tasks. These improvements stem from the rule-based Hit reward that directly optimizes for ranking accuracy, encouraging the model to assign higher probability mass to target itemic tokens. Notably, the Reco Reason task also benefits from RL training. This suggests that the refined ``recommendation intuition'' acquired through RL transfers to explanation generation, producing more coherent and relevant reasoning.

\begin{table}[h]
\caption{\textbf{Recommendation benchmark performance across post-training stages (Non-Thinking).}}
\label{tab:reco-posttrain-abliation}
\resizebox{\textwidth}{!}{
\begin{tabular}{l|cccccccc}
\hline
\textbf{Model}                                                      & \textbf{Video Rec}       & \textbf{Ad Rec}          & \textbf{Product Rec}        & \textbf{Label Cond.}     & \textbf{Interactive}     & \textbf{Label Pred.}     & \textbf{Item Understanding} & \textbf{Reco Reason}     \\ \hline
\textbf{Stage 1: Multi-task SFT }                       & 0.0324          & 0.0925          & 0.0532          & 0.0229          & \textbf{0.3461} & \textbf{0.6979} & 0.3274             & 3.8795          \\
\textbf{Stage 2:  On-Policy Distillation}  & 0.0304          & 0.0596          & 0.0330          & 0.0200          & 0.2419          & 0.6944          & \textbf{0.3319}    & 3.9479          \\
\textbf{Stage 3: Reinforcement Learning}                & \textbf{0.0370} & \textbf{0.0967} & \textbf{0.0536} & \textbf{0.0236} & 0.3458          & 0.6908          & 0.3209             & \textbf{4.0381} \\ \hline
\end{tabular}}
\end{table}

\section{Conclusion, Limitations, and Future Directions}
In this work, we presented \themodel, a comprehensive framework designed to bridge the gap between traditional recommendation systems and Large Language Models. We proposed \thebench, the first holistic benchmark for evaluating recommendation instruction-following capabilities, encompassing diverse tasks from fundamental prediction to complex reasoning. To facilitate reproducibility and scalable research, we open-sourced a full-stack training pipeline---including data processing, co-pretraining, and post-training protocols---and validated the scaling laws of recommendation capabilities. Extensive experiments demonstrate that our \themodel-Foundation models achieve state-of-the-art performance across \thebench and show exceptional transferability to external domains, proving the efficacy of our unified generative paradigm.
Despite these advancements, several limitations remain that point towards important future research directions. 

First, while our experiments confirm that the recommendation backbone significantly enhances downstream performance, the magnitude of these gains is currently constrained by tokenizer transferability. A promising avenue for future work lies in maximizing the reuse of foundation model priors while simultaneously ensuring high-quality item indexing (code quality) for downstream tasks. Second, maintaining the model's general intelligence and reasoning capabilities necessitates mixing vast amounts of general-domain text during training. Investigating the optimal data mixing ratios and improving data utilization efficiency are urgent challenges to balance domain-specific precision with general capabilities. Third, we observe that Chain-of-Thought reasoning currently yields improvements only in limited settings. This underscores the need for a more rigorous exploration of test-time scaling strategies to unlock consistent reasoning gains across diverse recommendation scenarios.

We believe that addressing these challenges requires collective effort from the community. By open-sourcing our entire framework, we invite contributions and encourage researchers to build upon \themodel. We hope this work serves as a solid foundation for future exploration, accelerating the evolution towards truly intelligent recommendation systems.
\newpage

\bibliographystyle{abbrvnat}
\nobibliography*
\bibliography{bibtex}

\clearpage

\appendix

\section{Contributions}
Within each role, authors are listed alphabetically by their first name. 

\begin{multicols}{3}
\noindent
\textbf{ Core Contributors} \\
 Guorui Zhou \\
 Honghui Bao \\
 Jiaming Huang\\
 Jiaxin Deng \\
 Jinghao Zhang \\
 Junda She \\
 Kuo Cai \\
 Lejian Ren \\
 Lu Ren \\
 Qiang Luo \\
 Qianqian Wang \\
 Qigen Hu \\
 Rongzhou Zhang \\
 Ruiming Tang \\
 Shiyao Wang \\
 Wuchao Li \\
 Xiangyu Wu \\
 Xinchen Luo \\
 Xingmei Wang \\
 Yifei Hu\\
 Yunfan Wu \\
 Zhanyu Liu \\
 Zhiyang Zhang \\
 Zixing Zhang

\noindent
\textbf{ Contributors} \\
 Bo Chen \\
 Bin Wen \\
 Chaoyi Ma \\
 Chengru Song \\
 Chenglong Chu \\
 Defu Lian \\
 Fan Yang \\
 Feng Jiang \\
 Hongtao Cheng \\
 Huanjie Wang \\
 Jiangxia Cao \\
 Kun Gai\\
 Pengfei Zheng \\
 Qiang Wang \\
 Rui Huang \\
 Siyang Mao \\
 Tingting Gao\\
 Wei Yuan \\
 Yan Wang\\
 Yang Zhou \\
 Yi Su \\
 Zexuan Cheng \\
 Zhixin Ling \\
 Ziming Li \\
\end{multicols}

\section{Appendix}
    \subsection{Evaluation Details for Item Understanding}
\label{appendix:eval_item_understanding}

For the Item Understanding task (Layer 0), we employ a rigorous \textbf{LLM-as-Judge} framework to evaluate the semantic alignment between the model-generated captions and the ground truth. The ground truth captions are generated by \textbf{Gemini-2.5-Pro} given the video frames and metadata. The evaluation process consists of three steps: Information Point Extraction, Semantic Matching, and Weighted Scoring. We use \textbf{Gemini-2.5-Flash-Lite} as the judge for extraction and matching.

\subsubsection{WIP Extraction and Matching}
We first prompt the LLM to decompose captions into atomic \textbf{Weighted Information Points (WIPs)}, each containing a fact statement and an importance score (1-5). Then, we employ the LLM to align model-generated WIPs with ground truth WIPs, identifying valid matches, hallucinations (unmatched model WIPs), and omissions (unmatched GT WIPs). The specific prompts are detailed in Section \ref{appendix:prompts_item_understanding}.

\subsubsection{Scoring Protocol}
To compute the final score, we calculate a \textbf{Double-Weighted F1 Score} that incorporates both the semantic importance of the information and the quality of the match. For each matched pair $(w_{gt}, w_{model})$, we compute a match quality score $q \in [0, 1]$, which is the \textbf{F1 score} calculated by \textbf{BERTScore} (using \texttt{bert-base-chinese}).

The contributions to True Positives ($TP_i$), False Negatives ($FN_i$), and False Positives ($FP_i$) for each sample $i$ are calculated as follows:
\begin{align*}
    TP_i &= \sum_{(w_{gt}, w_{model}) \in \text{Matches}} \text{score}(w_{gt}) \times q \\
    FN_i &= \sum_{(w_{gt}, w_{model}) \in \text{Matches}} \text{score}(w_{gt}) \times (1 - q) + \sum_{w_{gt} \in \text{Unmatched GT}} \text{score}(w_{gt}) \\
    FP_i &= \sum_{(w_{gt}, w_{model}) \in \text{Matches}} \text{score}(w_{model}) \times (1 - q) + \sum_{w_{model} \in \text{Unmatched Model}} \text{score}(w_{model})
\end{align*}

The F1 score for sample $i$ is computed as:
\begin{equation}
    F1_i = \frac{2 \cdot TP_i}{2 \cdot TP_i + FP_i + FN_i}
\end{equation}

The final score is the LLM-Judge Score, computed as the average of per-sample F1 scores:
\begin{equation}
    \text{LLM-Judge Score} = \frac{1}{N} \sum_{i=1}^{N} F1_i
\end{equation}

This metric penalizes both hallucinations (high FP) and omissions (high FN), while rewarding high-quality matches on important information points.

\subsection{Prompts for Item Understanding}
\label{appendix:prompts_item_understanding}

We provide the core instructions of the prompts used for the Item Understanding evaluation below.

\begin{demobox}[Prompt for WIP Extraction (Abridged)]
\footnotesize
\begin{CJK}{UTF8}{gbsn}
\textbf{任务}: 你是一位信息抽取专家。请将描述性文本分解为结构化的【原子化且唯一】的"信息点 (WIPs)"列表。

\textbf{输出结构}:
\begin{itemize}
    \item \texttt{info\_point}: 一个简洁的、陈述事实的短语。
    \item \texttt{importance\_score}: 整数 [1-5] (5=核心, 1=琐碎)。
\end{itemize}

\textbf{关键原则}:
\begin{enumerate}
    \item \textbf{原子性}: 每个点只包含一个独立事实。
    \item \textbf{唯一性}: 概念上不重复。
    \item \textbf{合并}: 将相似描述合并为一个代表性点。
\end{enumerate}

\textbf{打分指南}:
5 (灵魂/核心), 4 (骨架/关键事件), 3 (血肉/重要细节), 2 (背景/次要细节), 1 (琐碎信息)。
\end{CJK}
\end{demobox}

\begin{demobox}[Prompt for WIP Matching (Abridged)]
\footnotesize
\begin{CJK}{UTF8}{gbsn}
\textbf{任务}: 你是一位语义匹配专家。请对比"Ground Truth WIPs"和"Model-Generated WIPs"，识别匹配项、幻觉和漏报。

\textbf{匹配规则}:
\begin{enumerate}
    \item \textbf{语义核心}: 基于 \texttt{info\_point} 的核心含义进行匹配。
    \item \textbf{部分匹配}: 只要有语义重叠即可匹配 (例如："篮球比赛" 匹配 "球员打篮球")。
    \item \textbf{一对一}: 寻找最佳匹配对。
\end{enumerate}

\textbf{输出结构}:
\begin{itemize}
    \item \texttt{matches}: 语义相似的配对列表 \texttt{\{model\_wip, gt\_wip\}}。
    \item \texttt{unmatched\_model\_wips}: 幻觉 (False Positives)。
    \item \texttt{unmatched\_gt\_wips}: 漏报 (False Negatives)。
\end{itemize}
\end{CJK}
\end{demobox}
\subsection{Samples of Recommendation-Domain Pre-training Data}
\label{appendix:samples_rec_domain_pre_training}

In this section, we provide some simple samples for the three recommendation-domain pre-training data discussed in Section \ref{subsec:recommendation_data}.

\subsubsection{Itemic Dense Caption Data}

\begin{demobox}[Itemic Dense Caption Data]
    \footnotesize\raggedright
    \begin{CJK}{UTF8}{gbsn}
    视频\stok{item\_begin}\textcolor{tokenblue}{\texttt{<item\_a\_5028><item\_b\_6733><item\_c\_2559>}}\stok{item\_end} 展示了以下内容：视频内容聚焦在庆祝冬至这一重要节日的习俗，特别是享受饺子与汤圆等美食。视频表达了冬至节气的特色意义，以及人们对新一年开始的寓意。内容上，显现出浓浓的节日气氛与家庭温暖，可能会触动那些寻求传统节日体验和家的感觉的观众。视频还可能激发观众对中华传统文化的兴趣，以及对家人团聚时的美好记忆。通过美食与节日的结合，观众可感受到温馨和幸福，为冬至节日的到来营造了欢乐与期盼。\end{CJK}
    
    \vspace{0.3em}
    \rule{\linewidth}{0.4pt}
    \vspace{0.3em}
    
    \stok{item\_begin}\textcolor{tokenblue}{\texttt{<item\_a\_5028><item\_b\_6733><item\_c\_2559>}}\stok{item\_end} focuses on the customs of celebrating the Winter Solstice, a significant traditional festival, with a particular emphasis on enjoying delicacies such as dumplings and tangyuan. It conveys the unique cultural significance of the Winter Solstice solar term and its symbolism as the beginning of a new cycle. In terms of content, the video exudes a profound festive atmosphere and familial warmth, likely resonating with viewers seeking traditional holiday experiences and a sense of home. Furthermore, it may spark interest in traditional Chinese culture and evoke cherished memories of family reunions. By blending culinary traditions with the holiday spirit, the video allows the audience to experience warmth and happiness, fostering a sense of joy and anticipation for the arrival of the Winter Solstice.
\end{demobox}

\subsubsection{Sequential User Behavior Data}

\begin{demobox}[Sequential User Behavior Data]
    \footnotesize\raggedright
    \begin{CJK}{UTF8}{gbsn}
    用户的曝光序列为\mbox{\stok{item\_begin}\textcolor{tokenblue}{\texttt{<item\_a\_1023><item\_b\_5426><item\_c\_6422>}}\stok{item\_end}},\\\mbox{\stok{item\_begin}\textcolor{tokenblue}{\texttt{<item\_a\_3168><item\_b\_7950><item\_c\_4134>}}\stok{item\_end}},……；\\[2pt]
    其中长播列表是\mbox{\stok{item\_begin}\textcolor{tokenblue}{\texttt{<item\_a\_4988><item\_b\_7436><item\_c\_2477>}}\stok{item\_end}},\\\mbox{\stok{item\_begin}\textcolor{tokenblue}{\texttt{<item\_a\_5087><item\_b\_7888><item\_c\_4759>}}\stok{item\_end}},……；\\[2pt]
    点赞列表是\mbox{\stok{item\_begin}\textcolor{tokenblue}{\texttt{<item\_a\_3168><item\_b\_7950><item\_c\_4134>}}\stok{item\_end}},\\\mbox{\stok{item\_begin}\textcolor{tokenblue}{\texttt{<item\_a\_250><item\_b\_2310><item\_c\_4925>}}\stok{item\_end}},……
    \end{CJK}

    \vspace{0.3em}
    \rule{\linewidth}{0.4pt}
    \vspace{0.3em}
    
    The user's exposure sequence is \mbox{\stok{item\_begin}\textcolor{tokenblue}{\texttt{<item\_a\_1023><item\_b\_5426><item\_c\_6422>}}\stok{item\_end}},\\\mbox{\stok{item\_begin}\textcolor{tokenblue}{\texttt{<item\_a\_3168><item\_b\_7950><item\_c\_4134>}}\stok{item\_end}}, ...;\\[2pt]
    The long-viewed list is \mbox{\stok{item\_begin}\textcolor{tokenblue}{\texttt{<item\_a\_4988><item\_b\_7436><item\_c\_2477>}}\stok{item\_end}},\\\mbox{\stok{item\_begin}\textcolor{tokenblue}{\texttt{<item\_a\_5087><item\_b\_7888><item\_c\_4759>}}\stok{item\_end}}, ...;\\[2pt]
    The like list is \mbox{\stok{item\_begin}\textcolor{tokenblue}{\texttt{<item\_a\_3168><item\_b\_7950><item\_c\_4134>}}\stok{item\_end}},\\\mbox{\stok{item\_begin}\textcolor{tokenblue}{\texttt{<item\_a\_250><item\_b\_2310><item\_c\_4925>}}\stok{item\_end}}, ...
\end{demobox}

\subsubsection{Interleaved User Persona Grounding Data}

\begin{demobox}[Interleaved User Persona Grounding]
    \footnotesize\raggedright
    \begin{CJK}{UTF8}{gbsn}
    平台上有一名用户，她创作内容涵盖：8个其他，1个美食，1个数码，1个明星娱乐。
    
    她近期的搜索记录包括：怎么拍游戏视频、黑白头像可爱、……。
    
    她近期的购买记录包括：商品\stok{item\_begin}\textcolor{tokenblue}{\texttt{<item\_a\_6133><item\_b\_5060><item\_c\_5431>}}\stok{item\_end}，具体类型为【女装-裤子-休闲裤】，花费290元。
    
    她近期在视频\stok{item\_begin}\textcolor{tokenblue}{\texttt{<item\_a\_3316><item\_b\_7440><item\_c\_2022>}}\stok{item\_end}下评论了"这个短剧叫什么名字啊"；在视频\stok{item\_begin}\textcolor{tokenblue}{\texttt{<item\_a\_7822><item\_b\_1648><item\_c\_5756>}}\stok{item\_end}下评论了"嘻嘻嘻，真的吗？我也喜欢玩蛋仔派对，早就关注你了"；……。
    
    她点赞了视频\stok{item\_begin}\textcolor{tokenblue}{\texttt{<item\_a\_5743><item\_b\_930><item\_c\_1231>}}\stok{item\_end}……；
    
    收藏了视频\stok{item\_begin}\textcolor{tokenblue}{\texttt{<item\_a\_468><item\_b\_8186><item\_c\_5877>}}\stok{item\_end}……；
    
    分享了视频……。她关注的博主类型有：【其他】占47.58\%，【颜值】占16.52\%，【明星娱乐】占8.37\%，……。
    
    她近期观看的直播类型包括：【闲聊互动-热闹闲聊】分类下的直播点赞了6次，评论了59次；……
    
    她过去30天观看时间最长的1种短剧类型分别是:[解密\_悬疑]看了30.0分钟
    
    ……\end{CJK}
    \vspace{0.3em}
    \rule{\linewidth}{0.4pt}
    \vspace{0.3em}
    
      There is a user on the platform whose content creation covers: 8 in "Other," 1 in "Food," 1 in "Digital Tech," and 1 in "Celebrity \& Entertainment." 
      
      Her recent search history includes: "how to film game videos," "cute black and white avatars," ...
      
      Her recent purchase records include the item \stok{item\_begin}\textcolor{tokenblue}{\texttt{<item\_a\_6133><item\_b\_5060><item\_c\_5431>}}\stok{item\_end}, categorized as [Women's Wear - Pants - Casual Pants], costing 290 RMB.
      
      In terms of recent social interactions, she commented "What is the name of this short drama?" under the video \stok{item\_begin}\textcolor{tokenblue}{\texttt{<item\_a\_3316><item\_b\_7440><item\_c\_2022>}}\stok{item\_end}; and commented "Really? I love playing Eggy Party too! Actually, I've been following you for a while" under the video \stok{item\_begin}\textcolor{tokenblue}{\texttt{<item\_a\_7822><item\_b\_1648><item\_c\_5756>}}\stok{item\_end}; ...
      
      She also liked the video \stok{item\_begin}\textcolor{tokenblue}{\texttt{<item\_a\_5743><item\_b\_930><item\_c\_1231>}}\stok{item\_end} ...; 
      
      Bookmarked the video \stok{item\_begin}\textcolor{tokenblue}{\texttt{<item\_a\_468><item\_b\_8186><item\_c\_5877>}}\stok{item\_end} ...
      
      Shared the video ...
      
      The types of creators she follows are: [Others] accounting for 47.58\%, [Beauty] accounting for 16.52\%, [Entertainment] accounting for 8.37\%, and ...
      
      Her recent livestream viewing history includes: 6 likes and 59 comments in the [Chat \& Interaction - Lively Chat] category; and 9 comments in another...
      
      Over the past 30 days, her most-watched short drama genre was [Mystery/Suspense], with a total viewing time of 30.0 minutes...
\end{demobox}

\subsection{Detailed Data Composition and Token Budgets for Pre-training}
\label{appendix:data_composition_and_token_budgets}

Table \ref{tab:stage2-datasets} presents the main datasets used in Stage 2 of our Open model's pre-training. To ensure the model maintains robust general reasoning capabilities while adapting to recommendation tasks, we curate a diverse data mixture that is primarily composed of two parts:

\textbf{General Domain Data}: All general-domain datasets listed in the table are publicly available and can be downloaded from HuggingFace repository~\footnote{https://huggingface.co/datasets}. We incorporate a significant proportion of mathematical and coding datasets (e.g., Nemotron and OpenMath series) alongside general reasoning corpora to prevent the catastrophic forgetting of logic and instruction-following abilities.

\textbf{Recommendation Domain Data}: The three recommendation-domain datasets (Itemic Dense Caption Data, Interleaved User Persona Grounding Data, and Sequential User Behavior Data) are constructed from the raw meta data we have released, following the formats illustrated in the examples above (Section \ref{appendix:samples_rec_domain_pre_training}).

\begin{table}[htbp]
  \centering
  \caption{\textbf{Data mixture for Pre-training.} The table presents the distribution across general domains and recommendation domains, showing the sampling weight of each dataset and the subtotal ratio for each category.}
  \label{tab:stage2-datasets}
  \small
  \setlength{\tabcolsep}{3pt}
  \resizebox{0.95\textwidth}{!}{
  \begin{tabularx}{\textwidth}{lrc c c}
    \toprule
    \textbf{Dataset} & \textbf{Weight (\%)} & \textbf{Category} & \textbf{Subtotal (\%)} & \textbf{Token Budget} \\
    \midrule
    Nemotron\_CC\_Math\_v1\footnotemark & 37.4\% & Math & \multirow{12}{*}{62.34\%} & \multirow{12}{*}{29B} \\
    Nemotron\_Pretraining\_Code\_v1\footnotemark & 12.91\% & Code & & \\
    Nemotron\_CC\_v2\footnotemark & 5.59\% & Math, General, Code & & \\
    reasoning\_v1\_20m\footnotemark & 4.04\% & General & & \\
    OpenMathReasoning\footnotemark & 1.16\% & Math & & \\
    NuminaMath-QwQ-CoT-5M\footnotemark & 0.79\% & Math & & \\
    OpenCodeReasoning\footnotemark & 0.26\% & Code & & \\
    KodCode\_V1\_SFT\_R1\footnotemark & 0.09\% & Code & & \\
    Chinese-Reasoning-Distil-Data\footnotemark & 0.07\% & General & & \\
    medical-o1-reasoning-SFT\footnotemark & 0.02\% & Medical & & \\
    Bespoke-Stratos-17k\footnotemark & 0.01\% & Math, Code, Science & & \\
    \midrule
    Itemic Dense Caption Data & 36.94\% & Reco & \multirow{3}{*}{37.66\%} & \multirow{3}{*}{3B} \\
    Interleaved User Persona Grounding Data & 0.26\% & Reco & & \\
    Sequential User Behavior Data & 0.46\% & Reco & & \\
    \bottomrule
  \end{tabularx}
  }
\end{table}
\addtocounter{footnote}{-10}
\footnotetext{\url{https://huggingface.co/datasets/nvidia/Nemotron-CC-Math-v1}}
\stepcounter{footnote}\footnotetext{\url{https://huggingface.co/datasets/nvidia/Nemotron-Pretraining-Code-v1}}
\stepcounter{footnote}\footnotetext{\url{https://huggingface.co/datasets/nvidia/Nemotron-CC-v2}}
\stepcounter{footnote}\footnotetext{\url{https://huggingface.co/datasets/glaiveai/reasoning-v1-20m}}
\stepcounter{footnote}\footnotetext{\url{https://huggingface.co/datasets/nvidia/OpenMathReasoning}}
\stepcounter{footnote}\footnotetext{\url{https://huggingface.co/datasets/PrimeIntellect/NuminaMath-QwQ-CoT-5M}}
\stepcounter{footnote}\footnotetext{\url{https://huggingface.co/datasets/nvidia/OpenCodeReasoning}}
\stepcounter{footnote}\footnotetext{\url{https://huggingface.co/datasets/KodCode/KodCode-V1-SFT-R1}}
\stepcounter{footnote}\footnotetext{\url{https://huggingface.co/datasets/Mxode/Chinese-Reasoning-Distil-Data}}
\stepcounter{footnote}\footnotetext{\url{https://huggingface.co/datasets/FreedomIntelligence/medical-o1-reasoning-SFT}}
\stepcounter{footnote}\footnotetext{\url{https://huggingface.co/datasets/bespokelabs/Bespoke-Stratos-17k}}

\begin{table}[htbp]
\centering
\caption{\textbf{Data Composition and Token Budgets for Pre-training Stages.} This table illustrates the training configurations for the \textit{Open} and \textit{Pro} model variants across different stages, specifying the parameter focus, data domain distribution, and allocated token budgets.}
\label{tab:pretrain-token-composition}
\small
\setlength{\tabcolsep}{3pt}
\resizebox{0.95\textwidth}{!}{
\begin{tabular}{@{}llcccc@{}}
\toprule
\textbf{Model} & \textbf{Stage} & \textbf{Training} & \textbf{General-Domain} & \textbf{Reco-Domain} & \textbf{Token Budget} \\ \midrule
\multirow{2}{*}{OneRec} & stage1 & Itemic-related Parameters & \multirow{2}{*}{62.34\%} & \multirow{2}{*}{37.66\%} & 16B \\
                      & stage2 & Full-Parameter            &                          &                          & 32B \\ \midrule
\multirow{2}{*}{OneRec-Pro}  & stage1 & Itemic-related Parameters & 19\%    & 81\%    & 30B \\
                      & stage2 & Full-Parameter            & 53\%    & 47\%    & 130B \\ \bottomrule
\end{tabular}
}
\end{table}

For the OneRec variant, Stage 2 uses all datasets listed in Table \ref{tab:stage2-datasets}, totaling 32B tokens, while Stage 1 training data is sampled from this corpus. Table \ref{tab:pretrain-token-composition} details the comprehensive data composition and token budgets allocated for each pre-training stage, characterizing the strategic balance between general-domain and recommendation-domain knowledge. 

In contrast, the OneRec-Pro model leverages a significantly expanded budget of 130B tokens to achieve superior performance, by increasing the sampling ratios of the general-purpose datasets from Table \ref{tab:stage2-datasets}, integrating additional proprietary in-house datasets, and substantially expanding the scale of user-interaction data within the recommendation domain.

\subsection{Detailed Data Composition and Token Budgets for Multi-task SFT}
\label{appendix:data_composition_for_sft}

\begin{table}[htbp]
  \centering
  \caption{\textbf{Data Mixture for Multi-task SFT.} The table presents the distribution across reasoning and recommendation domains, showing the sampling weight of each dataset and the subtotal ratio for each category.}
  \label{tab:dataset-distribution}
  \small
  \setlength{\tabcolsep}{3pt}
  \begin{tabularx}{0.75\textwidth}{lcc r} 
    \toprule
    \textbf{Dataset} & \textbf{Weight (\%)} & \textbf{Category} & \textbf{Subtotal (\%)} \\
    \midrule
    OpenMathReasoning\footnotemark & 12.971\% & Math & \multirow{9}{*}{64.978\%} \\
    R1-Distill-SFT\footnotemark & 12.784\% & General & \\
    Infinity\_Instruct\footnotemark & 11.359\% & Instruction & \\
    OpenCoderReasoning\footnotemark & 11.130\% & Code & \\
    Chinese-Reasoning-Distil-Data\footnotemark & 4.552\% & General & \\
    Reasoning\_Multi\_subject\_RLVR\footnotemark & 4.376\% & Multi-subject & \\
    Reasoning\_KodCode\_V1\_SFT\_R1\footnotemark & 4.167\% & Code & \\
    DeepMath103K\footnotemark & 2.362\% & Math & \\
    medical-o1-reasoning-SFT\footnotemark & 1.277\% & Medical & \\
    \midrule
    Label prediction & 7.800\% & Reco & \multirow{8}{*}{35.022\%} \\
    SID to Caption generation & 7.493\% & Reco & \\
    Interactive recommendation & 6.392\% & Reco & \\
    Video recommendation & 3.971\% & Reco & \\
    Label conditional recommendation & 3.575\% & Reco & \\
    Product recommendation & 2.878\% & Reco & \\
    Ad recommendation & 2.820\% & Reco & \\
    Recommendation reasoning & 0.094\% & Reco & \\
    \bottomrule
  \end{tabularx}
\end{table}
\addtocounter{footnote}{-8}
\footnotetext{\url{https://huggingface.co/datasets/nvidia/OpenMathReasoning}}
\stepcounter{footnote}\footnotetext{\url{https://huggingface.co/datasets/ServiceNow-AI/R1-Distill-SFT}}
\stepcounter{footnote}\footnotetext{\url{https://huggingface.co/datasets/BAAI/Infinity-Instruct}}
\stepcounter{footnote}\footnotetext{\url{https://huggingface.co/datasets/nvidia/OpenCodeReasoning}}
\stepcounter{footnote}\footnotetext{\url{https://huggingface.co/datasets/Mxode/Chinese-Reasoning-Distil-Data}}
\stepcounter{footnote}\footnotetext{\url{https://huggingface.co/datasets/punwaiw/multi-subject-rlvr-final-reasoning-traces}}
\stepcounter{footnote}\footnotetext{\url{https://huggingface.co/datasets/KodCode/KodCode-V1-SFT-R1}}
\stepcounter{footnote}\footnotetext{\url{https://huggingface.co/datasets/zwhe99/DeepMath-103K}}
\stepcounter{footnote}\footnotetext{\url{https://huggingface.co/datasets/FreedomIntelligence/medical-o1-reasoning-SFT}}

\end{document}